\documentclass[aps,prd,twocolumn,preprintnumbers,superscriptaddress,nofootinbib,floatfix]{revtex4}

\pdfoutput=1

\usepackage{amssymb,amsmath}
\usepackage{mathtools}
\usepackage{bm}
\usepackage{graphicx}
\usepackage{epstopdf}
\usepackage{hyperref}
\usepackage{array}
\usepackage{soul}
\usepackage[usenames, dvipsnames]{color}
\usepackage{slashed}

\usepackage{soul,xcolor}

\setstcolor{blue}

\enlargethispage{\baselineskip}

\hypersetup{
     colorlinks   = true,
     citecolor    = blue
}

\def\ba{\begin{eqnarray}}
\def\ea{\end{eqnarray}}
\def\be{\begin{equation}}
\def\ee{\end{equation}}

\begin{document}

\newcommand*{\bfrac}[2]{\genfrac{}{}{0pt}{}{#1}{#2}}
\newcommand{\tos}{t_{\rm osc}}
\newcommand{\Ho}{H_{\rm osc}}
\newcommand{\ho}{h_{\rm osc}}
\newcommand{\co}{\chi_{\rm osc}}
\newcommand{\mP}{m_{\rm Pl}}
\newcommand{\THi}{\tau_{\rm Higgs}}
\newcommand{\THu}{\tau_{\rm Hubble}}
\newcommand{\TRH}{T_{\rm reh}}
\newcommand{\hh}{\langle h^2\rangle}
\newcommand{\LL}{\mathcal{L}}
\newcommand{\WW}{\mathcal{W}_k}
\newcommand{\HH}{\mathcal{H}}
\newcommand{\hrms}{h_{\rm rms}}
\newcommand{\td}{t_{\rm reh}}
\newcommand{\tdo}{t_{\rm reh,0}}
\newcommand{\Nr}{N_{\rm reh}}

\newcommand{\es}[1]{ {[{\bf ES}: #1]}}
\newcommand{\ps}[1]{ {[{\bf PS}: #1]}}
\newcommand{\lv}[1]{\textcolor{RawSienna}{[{\bf LV}: #1]}}
\newcommand{\edit}[1]{\textcolor{orange}{ #1}}
\newcommand{\esedit}[1]{\textcolor{cyan}{ #1}}

\newcommand{\FIRSTAFF}{\affiliation{Department of Physics,
		University of Michigan,
		Ann Arbor, MI 48109, USA}}
\newcommand{\SECONDAFF}{\affiliation{The Oskar Klein Centre for Cosmoparticle Physics,
	Department of Physics,
	Stockholm University,
	AlbaNova,
	10691 Stockholm,
	Sweden}}
\newcommand{\THIRDAFF}{\affiliation{Nordita,
	KTH Royal Institute of Technology and Stockholm University
	Roslagstullsbacken 23,
	10691 Stockholm,
	Sweden}}
\newcommand{\FOURTHAFF}{\affiliation{Department of Physics,
	University of Illinois at Urbana-Champaign,
	Urbana, IL 61801,
	USA}}
\newcommand{\FIFTHAFF}{\affiliation{Nikhef,
Science Park 105,
1098 XG Amsterdam, The Netherlands}}	

\newcommand{\SIXTHAFF}{\affiliation{Institute Lorentz of Theoretical Physics, 
University of Leiden, 2333CA Leiden, The Netherlands}}

\renewcommand\({\left(}
\renewcommand\){\right)}
\renewcommand\[{\left[}
\renewcommand\]{\right]}

\title{The Higgs Boson can delay Reheating after Inflation }

\author{Katherine Freese}
\email[Electronic address:]{ktfreese@umich.edu}
\FIRSTAFF
\SECONDAFF
\THIRDAFF

\author{Evangelos I. Sfakianakis}
\email[Electronic address:]{e.sfakianakis@nikhef.nl}
\FOURTHAFF
\FIFTHAFF
\SIXTHAFF

\author{Patrick Stengel}
\email[Electronic address:]{patrick.stengel@fysik.su.se}
\FIRSTAFF
\SECONDAFF

\author{Luca Visinelli}
\email[Electronic address:]{luca.visinelli@fysik.su.se}
\SECONDAFF
\THIRDAFF

\preprint{ NORDITA-2017-120}
\preprint{LCTP-17-10}
\preprint{Nikhef 2017-051}

\date{\today}

\begin{abstract}

The Standard Model Higgs boson, which has previously been shown to develop an effective vacuum expectation value during inflation, can give rise to large particle masses during inflation and reheating, leading to temporary blocking of the reheating process and  a lower reheat temperature after inflation.
  We study the effects on the multiple stages of reheating: resonant particle production (preheating) as well as perturbative decays from coherent oscillations of the inflaton field. 
  Specifically,  we study both the cases of the inflaton coupling to Standard Model fermions through Yukawa interactions as well as to Abelian gauge fields through a Chern-Simons term.  We find that, in the case of perturbative inflaton decay to SM fermions, reheating can be delayed due to Higgs blocking and the reheat temperature can decrease by up to an order of magnitude.
In the case of gauge-reheating, Higgs-generated masses of the gauge fields can  suppress preheating even for large inflaton-gauge couplings. In extreme cases, preheating can be shut down completely and must be substituted by perturbative decay as the dominant reheating channel. Finally, we discuss the distribution of reheat temperatures in different Hubble patches, arising from the stochastic nature of the Higgs VEV during inflation and its implications for the generation of both adiabatic and isocurvature fluctuations.

\end{abstract}
\maketitle

\section{Introduction}
\label{sec:Introduction}

Inflation was first proposed by Guth in Ref.~\cite{Guth:1980zm} to solve several cosmological puzzles that plagued the old Big Bang Theory: an early period of accelerated expansion explains the homogeneity, isotropy, and flatness of the Universe, as well as the lack of relic monopoles.  Subsequently, Linde~\cite{Linde:1981mu}, as well as Albrecht and Steinhardt~\cite{Albrecht:1982wi} suggested rolling scalar fields as a mechanism to drive the dynamics of inflation.  In these models, quantum fluctuations of the inflaton field generate density fluctuations that leave imprints on the cosmic microwave background (CMB) and lead to the formation of the large scale structure in the Universe, i.e. the galaxies and clusters we observe today. After driving the exponential expansion of the Universe, inflation must also successfully reheat the Universe to ensure the transition to a radiation dominated state before Big Bang Nucleosynthesis (BBN). In models with rolling scalar fields, the transfer of energy from the inflaton to relativistic particles can be facilitated by couplings of the inflaton to Standard Model (SM) fields\footnote{There is also the possibility of the inflaton transferring its energy to one or more intermediary fields, which then couple to SM particles. We will not delve into such models here, due to their inherent ambiguity.}.
 While the details of such couplings are highly model dependent, in general, the associated reheating can occur through the perturbative decay of the inflaton or through resonant particle production (preheating). Although the decay products of the inflaton are usually considered to be massless, recent work suggests that SM fields can acquire large masses due to the Higgs boson condensate which develops during inflation~\cite{Enqvist:2013kaa}.  In this work we examine the delay of the reheating process due to large SM masses during and after inflation and the possible implications of inhomogeneities arising from the stochastic motion of the Higgs field during inflation.

Enqvist et al.~\cite{Enqvist:2013kaa} showed that during inflation, the SM Higgs boson can develop a mass and electroweak (EW) symmetry can be treated as effectively broken~\cite{Kusenko:2014lra}. The effective Higgs mass arises due to quantum fluctuations of the Higgs field --- similar to the quantum fluctuations of the inflaton field that generate the density fluctuations responsible for large scale structure~\cite{Starobinsky:1994bd}.  The expectation value of the Higgs amplitude over the entire inflating patch is vanishing $\langle h_I \rangle = 0$ due to the symmetric potential (where subscript $I$ indicates initial value at the onset of inflaton oscillations). However, the variance is non-zero and the typical Higgs amplitude (the effective Vacuum Expectation Value or VEV for short) in a random Hubble patch at the end of inflation is given by a root mean square value $h_I = \sqrt{\hh} \propto H_I$ where $H_I$ is the Hubble scale at the end of inflation. The effective nonzero Higgs VEV during inflation then gives mass to all SM particles that couple to the Higgs. For reheating to occur, the inflaton must decay to other particles; yet this decay may be blocked if the inflaton mass is lower than the mass of decay products induced by the effective nonzero Higgs VEV $h_I$. In short, in any model where the inflaton decays to SM particles which are coupled to the Higgs, reheating can be delayed until the Hubble-valued Higgs condensate dissolves. The same analysis as the one presented here holds in principle if the inflaton is coupled to a similarly Higgsed Dark sector, although the quantitative details can differ.
 
The lower reheating temperature that results due to the nonzero Higgs mass during inflation has various consequences. In the approximation that the Higgs condensate acquires a space-independent VEV, a constant suppression of the reheat temperature occurs across the observable Universe (however, see below for discussion of inhomogeneties). A global suppression of the reheat temperature may affect constraints on inflation models arising from the CMB which are often portrayed in the ``$n_s - r$" plane.  Here, $n_s$ refers to the spectral index of the density perturbations produced by the model and $r$ is the tensor-to-scalar ratio, i.e. the ratio of the amplitude of gravitational waves to density perturbations. Predictions of specific inflation models may be visually compared to data from CMB experiments by plotting both the theoretical predictions and the data in the $n_s - r$ plane. However, the location of the predictions in the plane depends on the reheat temperature~\cite{Cook:2015vqa, Liddle:2003as, Martin:2006rs, Lorenz:2007ze, Adshead:2010mc, Easther:2011yq, Dai:2014jja, Martin:2014nya, Drewes:2015coa, Feng:2003nt}. The delayed reheating we find in this paper generally shifts predictions towards lower values of $n_s$ and larger values of $r$. Since current constraints on inflationary models consider scenarios in which reheating occurs at an unspecified point between the end of inflation and BBN, our work does not aim to extend these bounds, but rather to shrink them by providing a more accurate description of (p)reheating, given a specific model.

Another consequence of the lower reheat temperature is the possible effect on baryogenesis. Several models have been proposed that tie the origin of the matter/antimatter asymmetry of the Universe to inflation. In these models a lepton (helicity) asymmetry is produced at the end of inflation and is later converted to a baryon asymmetry through the electroweak sphaleron process~\cite{Giudice:1999fb, Adshead:2015jza, Kusenko:2014lra, Pearce:2015nga, Yang:2015ida, Alexander:2004us, Caldwell:2017chz}.  In these models, if the reheat temperature is too high, the lepton asymmetry generated during reheating could be suppressed by rapid lepton number violating interactions~\cite{Yang:2015ida, Adshead:2017znw}. A lower reheat temperature could avoid such processes. 

However, the stochastic evolution of the Higgs field during inflation will inevitably lead to an inhomogeneous delay of the reheating process, hence the Universe will be comprised at the end of inflation by a collection of Hubble patches with different reheat temperatures. This can lead to both adiabatic and (in some leptogenesis models) baryon isocurvature perturbations. The study of fluctuations arising due to the stochasticity of the Higgs VEV opens up a new avenue for analyzing and constraining inflationary models or mapping the Higgs potential at high energies.
We will refrain from a detailed discussion of these aspects until section\ \ref{sec:discussion}.
 
Reheating can occur in multiple stages, including preheating (resonant particle production), followed by perturbative decays (from coherent oscillations of the inflaton field)\footnote{Depending on the specific model, each of these two stages can be subdominant or even absent.}. In this paper, we assume that the inflaton couples primarily to SM  particles that develop masses when the Higgs field acquires a mass. Instead, if the inflaton were to decay to a massless gauge mode in the broken phase (the analog of the photon at lower temperatures) or to neutrinos, then the effective Higgs mass during inflation would not in any way affect reheating, since the Higgs does not provide a mass to these particles\footnote{Majorana neutrinos that acquire their mass due to the see-saw mechanism are also Higgsed~\cite{Adshead:2015jza}. However, since the nature of neutrinos is uncertain at present, we do not try to provide any detailed treatment here.}. We note that the direction of the Higgs VEV may or may not coincide with the current direction of Spontaneous Symmetry Breaking, hence the massless direction during inflation is not in general today's photon. We will examine in detail the various types of reheating of the inflaton, given the nonzero Higgs VEV and resultant SM particle masses during inflation. 

Perturbative Decay:  
After inflation the inflaton field can decay perturbatively into light particles forming a thermal bath, assuming the requisite coupling of the inflaton to SM fields.
 While often not as efficient at draining power from the inflaton condensate as resonant particle production, a strong enough interaction between the inflaton and the SM particles can ensure that the Universe be radiation dominated before BBN.  We specifically calculate the inflaton decay to fermions, which develop a mass due to the effective nonzero Higgs VEV during inflation. For simplicity, we only consider the case of a Yukawa interaction of the inflaton with an arbitrary SM fermion of undetermined mass today (after the usual EW phase transition), such that we can calculate the perturbative decay width of the inflaton. We are thus able to determine, in general, the range of SM Yukawa couplings, for a given inflationary energy scale, for which the effects of the Higgs condensate on perturbative reheating are significant. 
Compared to the masses SM fields receive from the effective Higgs VEV which develops during inflation (with the VEV set by the inflation scale), the masses from the usual EW symmetry breaking are negligible, provided that the inflation scale is not too low.
Thus  the results of our calculations of the perturbative width should remain essentially the same for any SM final state particles  (as long as the final state particles become massive due to the Higgs mechanism); the details of the phase space of the decay products would vary for different inflaton-SM interaction terms and particular final states but would not substantially change our final results.

Non-perturbative Resonant Particle Production: The first type of reheating to typically take place, and frequently the most effective one, is non-perturbative preheating, which leads to the resonant production of  bosons. There are two possible contributions to preheating:  tachyonic resonance (where the effective inflaton oscillation frequency-squared first crosses zero with $\omega^2<0$), followed by parametric resonance.  To reduce complications, we will only consider a massive $U(1)$ field during inflation, as a proxy for the massive electroweak gauge bosons during (and immediately after) inflation. To be more concrete, we study the case of natural inflation~\cite{Freese:1990rb} 
for our investigation of non-perturbative decay. Here, the inflaton is an axion, designed to avoid the fine-tuning that plagues most other models of inflation.  A shift symmetry provides the flat potential required for a successful inflation model~\cite{Adams:1990pn}.

The coupling of an axion-like inflaton to a $U(1)$ gauge boson has been shown to allow for complete preheating, in which the entirety of the inflaton's energy density is transferred to the gauge fields within a single inflaton oscillation~\cite{Adshead:2015pva}. In the particular case where the $U(1)$ gauge field is identified as the hypercharge sector of the SM, this very efficient energy transfer can lead to the generation of large-scale magnetic fields with possible cosmological relevance for the explanation of the observed Blazar spectra~\cite{Adshead:2016iae, Neronov:1900zz, Tavecchio:2010mk,Dolag:2010ni, Essey:2010nd, Taylor:2011bn, Takahashi:2013lba, Finke:2013tyq, Kachelriess:2012mc}. Although our calculations specifically assume a derivative coupling of the inflaton to gauge bosons which arises in models of natural inflation~\cite{Adams:1992bn, Gaillard:1995az,Kawasaki:2000ws,Banks:2003sx,Hsu:2003cy,Hsu:2004hi,Freese:2004un,Kim:2004rp}, the results of this work can be easily generalized to any model in which preheating occurs through resonant massive boson production.  We also discuss perturbative decays of the inflaton field in this case of axion-like couplings when resonant particle production is completely blocked by the induced mass of the final state bosons. 

The outline of this work is as follows. In section~\ref{sec:HiggsCondensate}, we review the dynamics of the Higgs condensate during and after inflation. In section~\ref{sec:perturbative}, we investigate the effects of SM particle masses on the perturbative decay of the inflaton. In section~\ref{sec:gauge}, we consider resonant production of massive gauge bosons during preheating. We summarize our conclusions and consider further applications of this work in section~\ref{sec:discussion}. 

\section{Higgs Condensate} \label{sec:HiggsCondensate}

 We consider the Higgs doublet and its potential in the form
\ba
	\Phi &=& \frac{1}{\sqrt{2}}\left(\bfrac{0}{h}\right),\\
	V_H(h) &=& \frac{\lambda}{4}\left(\Phi^\dag\Phi - \frac{\nu^2}{2}\right)^2 \approx \frac{\lambda}{4}\,h^4, \label{eq:higgs_potential}
\ea    
where $\nu = 246\,$GeV and $\lambda = \lambda(\mu)$ is the Higgs quartic self-coupling, whose running depends on the renormalization parameter $\mu$. In the absence of new physics, the quartic coupling constant $\lambda(\mu)$ is known up to computations involving three-loop diagram corrections~\cite{Degrassi:2012ry, buttazzo2013}. The coupling is positive up to an instability scale $\mu_{\rm inst}$ for which $\lambda(\mu_{\rm inst}) = 0$; above the instability scale, the Higgs quartic coupling becomes negative and the Higgs field slides to large values. Since the instability scale is computed to be $\mu_{\rm inst} \approx 10^{11}\,$GeV~\cite{Degrassi:2012ry, buttazzo2013, bezrukov2012}, while the inflation scale $H_I$ can be as high as $10^{14}\,$GeV, in principle we would require new physics in order to stabilize the Higgs vacuum at the inflation scale. This problem is particularly persistent since, in order to minimize radiative corrections to the Higgs potential, we set the value of the renormalization scale $\mu \sim h_I$, where the amplitude of the Higgs field during inflation lies below the scale of inflation, $h_I \lesssim H_I$. However, the computation of the running of the coupling constant strongly depends on the value of the mass of the top quark, to the extent that even a slight deviation from the central value used has large impacts on the estimation of the energy scale $\mu_{\rm inst}$. In particular, when the value of the top quark mass is lowered by three $\sigma$ from the central measured value, we push $\mu_{\rm inst} \sim 10^{15}\,$GeV~\cite{Degrassi:2012ry, buttazzo2013, bezrukov2012, Enqvist:2014bua}. With these assumptions, we estimate the lowest value for the Higgs quartic self-coupling as $\lambda_I \sim 10^{-3}$ at the inflationary scale~\cite{Degrassi:2012ry}. We use this value of $\lambda$ as our canonical value when estimating numbers throughout the text, but stress that this value is uncertain and heavily depends on the value of the SM parameters as well as inputs from additional new physics beyond the SM. Furthermore, we are not including possible contributions from the generation of $\langle h^2 \rangle$ due to the matter-dominated expansion of the background space-time, as discussed in Ref.~\cite{Markkanen:2017edu}.

The SM Higgs boson that is minimally coupled to gravity  is a light spectator field during inflation.  Though initially rolling classically, the Higgs field soon reaches a regime dominated by quantum fluctuations. To set the initial conditions for the Higgs field at the end of inflation (and the start of the inflaton oscillations leading to the reheating or preheating stages), we use the fact that the superhorizon modes of the Higgs field follow a random walk during the final stages of inflation, with the probability distribution function (PDF) given by~\cite{Starobinsky:1994bd}
 \be
 	f_{\rm eq}(h) = \(\frac{32\pi^2\lambda_I}{3}\)^{1/4}\frac{1}{\Gamma\(\frac{1}{4}\)H_I}\exp\(-\frac{2\pi^2\lambda_I h^4}{3H_I^4}\),
	\label{eq:pdfhiggs}
\ee
where $H_I$ and $\lambda_I$ are the Hubble rate and the Higgs quartic self-coupling evaluated at the end of inflation, $\Gamma(x)$ is the Gamma function and $\Gamma(1/4)\simeq 3.625$. Throughout the paper, the subscript $I$ refers to the initial time of onset of inflaton oscillations (referred to as the end of inflation). It is straightforward to check that the above PDF is properly normalized as $\int_{-\infty}^{\infty} f_{\rm eq}(h)  dh=1$.
The Higgs field is thus distributed in different Hubble patches according to the above PDF, so that we can take the Higgs VEV in each Hubble patch to be constant and have a magnitude $h$ with a probability given by  Eq.~\eqref{eq:pdfhiggs}. The dispersion of the PDF is
\be
	\hh = \int dh f_{\rm eq}(h) h^2 ,
\ee
 yielding an effective VEV
\ba
	 \sqrt{\hh} = 0.36\lambda_I^{-1/4}H_I\, ,
	\label{eq:centralvalue}
\ea
which gives masses to the Higgs and other SM particles. We take $\sqrt{\hh}$ to be the initial value of the Higgs field at the end of inflation, $h_I$. We note that $h_I \gg \nu$~\cite{Enqvist:2014tta}.  

 {It is important to keep in mind that the Higgs field displacement described in Eq.~\eqref{eq:centralvalue} is only true if one considers the de-Sitter equilibrium solution. However, inflation is not an infinitely long de-Sitter period.  The deviation from a de-Sitter universe becomes more pronounced towards the end of inflation, where the rate of change of the Hubble scale, described by the slow-roll parameter $\epsilon \equiv -\dot H/H^2$ grows towards unity. Hence, one can expect that the true distribution of Higgs field at the end of inflation is not characterized by the Hubble scale at the end of inflation, but rather at the Hubble scale at some previous time, when the de-Sitter solution was still a good approximation. A thorough investigation of the stochastic evolution of spectator fields was given in Ref.~\cite{Hardwick:2017fjo}, where it was found that in some cases spectator field displacements may be larger than the ones suggested by the de-Sitter equilibrium solution. Since we do not wish to digress into a case-by-case examination of detailed spectator field dynamics, we will use Eq.~\eqref{eq:centralvalue}, keeping in mind that  it provides a --sometimes conservative-- estimate of the true spread of Higgs values at the end of inflation.}

Once the inflaton begins to oscillate, the Universe is matter dominated and the Hubble constant drops as $H \sim a^{-3/2}$.
The dynamical evolution of the Higgs field may be described by the equation of motion for a scalar field in a quartic potential,
\be
	\ddot{h} + 3H\dot{h} + \lambda h^3 = 0,
	\label{eq:Higgs}
\ee
and we define the effective Higgs mass squared as
\ba
	m_h^2 \equiv 3\lambda  h^2  \, .
\ea
Eq.~\eqref{eq:Higgs} neglects interactions with other fields; the role of coupling to gauge fields will be discussed shortly in Eq.~\eqref{eq:higgslagrangian} and the subsequent discussion.

Since the self-interaction term in Eq.~\eqref{eq:Higgs} is negligible immediately after the the end of inflation, the amplitude of the Higgs field remains ``frozen" at the value it had at the end of inflation $h_I$, until it starts oscillating. Using Eq.~\eqref{eq:centralvalue}, we obtain the effective Higgs mass squared from the end of inflation until the onset of Higgs oscillations,
\ba
	m_{h_I}^2   \sim  0.4 \lambda_I^{1/2} H_I^2.
\ea
Oscillations of the Higgs field begin once the Hubble constant has dropped to the value $\Ho \approx m_h$, that is when the mass term overcomes the friction term in Eq.~\eqref{eq:Higgs}, at which point the value of the Higgs field is
\ba
	\ho \sim h_I \, ,
	\label{eq:higgsosc}
\ea
where the subscript ``osc'' indicates the time when the Higgs oscillations begin and we have assumed there is very little running of $\lambda$ between the end of inflation and the onset of Higgs oscillations. The ratio between the Hubble constants at these two times is \footnote{This  equation differs from Eq.~(2.6) in Ref.~\cite{Enqvist:2013kaa}, which was derived using the assumption that $h \propto a^{-1}$ even for $t < \tos$, an assumption that does not apply during this time period.\label{footnote_Enqvist}}
\ba
	\frac{\Ho}{H_I} \sim 0.6 \lambda_I^{1/4} \,.
\ea


For $\lambda_I \sim 10^{-3}$, the Higgs field begins to oscillate about five Hubble times after the end of inflation~\cite{Enqvist:2013kaa} (as we show in Fig.~\ref{fig:higgs} below, this corresponds to a few $e$-folds). As we will see below, the time $\tos$ approximately coincides with the time at which the transition between the two contributions to inflaton reheating occurs. Preheating (both tachyonic and parametric resonance) takes place before $\tos$ while perturbative decay begins after $\tos$. As a consequence, the Higgs field varies slowly during preheating and can be taken to be constant, while a detailed treatment of its evolution is relevant for 
perturbative inflaton decays due to the oscillatory behavior of the Higgs field after $\tos$.

As discussed in Ref.~\cite{Enqvist:2013kaa}, the energy density in the Higgs field is eventually dissipated through resonant production of W bosons via the dominant decay mode $h\to WW$. The dynamics of the Higgs field doublet coupled to the gauge field $W^a_\mu$ is described by the Lagrangian term
\be
	\LL_{\Phi+W} = (D_\mu\Phi)^\dag D^\mu\Phi - V_H(\Phi) + \frac{1}{4}G_a^{\mu\nu}G^a_{\mu\nu},
	\label{eq:higgslagrangian}
\ee
where the covariant derivative of the Higgs to the gauge field is $D_\mu\Phi = \left(\partial_\mu - ig\tau^a W^a_\mu\right)\Phi$, $g$ is a coupling constant, $\tau^a$ is a set of generators for the gauge group, and $G_a^{\mu\nu}$ is the field strength, defined through $G^a_{\mu\nu} = \partial_\mu W^a_\nu -  \partial_\nu W^a_\mu + g\epsilon^{abc}W^b_\mu W^c_\nu$. Eq.~\eqref{eq:Higgs} for the dynamical evolution of the Higgs field can be obtained from this Lagrangian by  neglecting the gauge couplings of the Higgs, a reasonable approximation throughout most of the reheating process  prior to the final dissipation of the Higgs into gauge bosons. As we will show in the case of perturbative inflaton decay, for $t\gtrsim \tos$, the Higgs field redshifts significantly as $h \sim a^{-1}$ before backreaction effects become important. The decay time of the Higgs condensate can then be reasonably approximated by the time at which the backreaction term in the equation of motion becomes comparable to the co-moving amplitude of the Higgs field. With this approximation in mind, we write the equation of motion~\cite{Enqvist:2013kaa} for the gauge field~\footnote{We ignore the non-Abelian self-interactions of the gauge fields, which may change the Higgs condensate decay time somewhat, but should not drastically affect our overall results. }, 
\be
	\ddot{\WW} +\omega^2_k\WW = 0,
	\label{eq:gauge}
\ee
where $W_k$ is the Fourier transform of the transverse component of the gauge field in Eq.~\eqref{eq:higgslagrangian} and $\WW = a^{3/2}W_k$, while the time-dependent frequency of the mode with wavenumber $k$ is
\be
	\omega^2_k = \frac{k^2}{a^2} + \frac{g^2h^2}{4} - \frac{3}{2}\frac{\ddot{a}}{a} - \frac{3}{4}H^2.
	\label{eq:motion_Wboson}
\ee
Note that the sum of the last two terms in the expression for $\omega_k^2$ vanishes in a matter-dominated background. The corresponding occupation number is given by~\cite{Greene:1997fu}
\be
	n_k = {1 \over {2\omega_k}} \left({| \dot{\WW} |^2} + \omega_k^2 | \WW |^2 \right) - {1 \over 2},
	\label{eq:numberdensityW}
\ee
from which we calculate the effective Higgs mass term induced by W-bosons using an approximate expression for the expectation value $\langle W^2 \rangle$,
\be
	m_{h (W)}^2 \simeq {g^2 \over 4} \int \frac{d^3 k}{(2\pi a)^3} {n_k \over \omega_k } .
	\label{eq:mass_Wboson}
\ee
Assuming that the dominant decay mode of the Higgs condensate is non-perturbative W-boson production, the condensate decays approximately when $m_{h (W)}^2$ has grown to reach the value 
\be
\label{eq:hit}
	m_{h (W)}^2 \simeq m_h^2 \,\,\,\,\,\,\, ( {\rm condition} \,\, {\rm for} \,\, {\rm Higgs} \,\, {\rm decay}) .
\ee
This condition is an approximation for the conservation of energy density stored in the Higgs condensate. Since the dominant decay channel for the Higgs condensate is resonant W-boson production, then, if the condition in Eq.~\eqref{eq:hit} is met, the energy density implied by the effective mass of the Higgs field, $m_h^2$, will have been depleted.
As we show in the next section, the details of the condensate decay are only relevant in very specific cases of perturbative reheating.

We note that, for values of the Hubble scale during inflation that are much smaller than what proposed above, the delayed reheating effects considered in this paper become unimportant. The effective Higgs VEV in Eq.~\eqref{eq:centralvalue} becomes smaller both because of the smaller value of $H_I$ and also because $\lambda_I$ runs to higher values.  The resulting fermion masses due to the Higgs VEV are then also smaller and no longer important.

\section{Perturbative Inflaton Decay} \label{sec:perturbative}


\subsection{Neglecting backreaction}

If we assume the inflaton has begun to oscillate as a massive scalar field by the end of inflation, under the influence of a quadratic potential $V_\phi = m_\phi^2 \phi^2 /2$, the set of equations which describes the perturbative reheating is
\ba
	\dot{\rho}_\phi +3H\rho_\phi &=& -\Gamma_\phi \rho_\phi, \label{eq:inflaton_res}\\
	\dot{\rho}_R + 4H\rho_R &=& \Gamma_\phi \rho_\phi,\label{eq:radiation_res}\\
	H^2 = \(\frac{\dot{a}}{a}\)^2 &=& \frac{8\pi}{3}\(\rho_\phi + \rho_R\),\label{hubblerate}
\ea
where $\rho_\phi$ is the energy density of the inflaton and $\rho_R$ is the energy density of radiation resulting from the decay of the inflaton with a decay rate $\Gamma_\phi$. For simplicity, we assume the inflaton has a Yukawa coupling to SM fermions, yielding
\ba
	\Gamma_\phi = \Gamma_0 \left(1 - {4 m_f^2 \over m_\phi^2} \right)^{3/2} \Theta \left(m_\phi^2 - 4 m_f^2 \right) ,
	\label{eq:gamma_higgs}
\ea
where $\Gamma_0$ is the decay width in the massless fermion limit and 
\ba
	m_f^2 = {1\over 2}y^2  h^2  ,
	\label{eq:fermion_mass}
\ea
is the effective fermion mass induced by the Higgs condensate with Yukawa coupling $y$. In Eq.~\eqref{eq:gamma_higgs}, we have included the $\Theta$-function (step-function) to model the phase-space blocking due to large effective fermion masses. As mentioned in section~\ref{sec:Introduction}, the results of our calculations of the perturbative width should remain essentially the same for any SM final particles (as long as they become massive due to the Higgs); the details of the phase space of the decay products would vary for different inflaton-SM interactions or choice of final states but would not substantially change our basic results. 

The decay of the inflaton field is controlled by the decay rate in Eq.~\eqref{eq:gamma_higgs} and it is thus negligible (blocked) when 
$4m_f^2 > m_\phi^2$, i.e. when
\be
	\frac{h^2}{m_\phi^2} > \frac{1}{2y^2}.
	\label{eq:blocking}
\ee
Fig.~\ref{fig:higgs} shows the value of the Higgs field squared amplitude in units of $m_\phi^2$ (black solid line), as a function of the number
of $e$-folds after the inflaton starts oscillating, for the choice $H_I = m_\phi/2$ and $\lambda_I =10^{-3}$. We also plot the quantity $1/2y^2$ for $y = 1$ (green dotted line), $y = 10$ (red dot-dashed line), and $y = 100$ (blue dashed line). For a given value of $y$, whenever $h$ falls below the value given by the blocking condition in Eq.~\eqref{eq:blocking}, the inflaton field decays in a series of burst events. This picture is valid whenever the period of the oscillations in the Higgs field $\THi$ is much larger than the characteristic Hubble timescale {$ {\THu} = 1/H$. In the opposite limit, $ {\THi} \ll  {\THu}$, the blocking in Eq.~\eqref{eq:blocking} still applies provided that the Higgs field is replaced with its root-mean-squared value $\hrms(t)$, shown with the black dashed line in Fig.~\ref{fig:higgs}\footnote{We wish to distinguish the initial condition average $h_I$ (that was defined in terms of the stochastic superhorizon probability distribution) from the root-mean-square  which is obtained by $\hrms \sim \rho_h^{1/4}$ where $\rho_h = \dot{h}^2/2 + V_H(h)$.}. In this second scenario, the decay does not occur in bursts since the oscillations in $\hrms(t)$ are suppressed. Unfortunately the true oscillations of the Higgs field are likely to be in an intermediate regime where the period is comparable to the Hubble rate, $ {\THi} \simeq  {\THu}$, in which case the usual calculation of a perturbative decay width is not appropriate since the mass of the final states is varying too slowly to yield a root-mean-squared value and too quickly to resolve the times when Eq.~\eqref{eq:blocking} is satisfied. Thus, we present our results assuming the two limiting cases in which we are confident of the perturbative width calculation, and we leave a more realistic calculation to future work. The two cases split when the Higgs starts oscillating\footnote{Before $\tos$, non-perturbative particle production described in the next section can be much more efficient in reheating than any perturbative particle production described in this section.} (i.e. at $t_{\rm osc})$.  
\begin{figure}
\begin{center}
	\includegraphics[width=\linewidth]{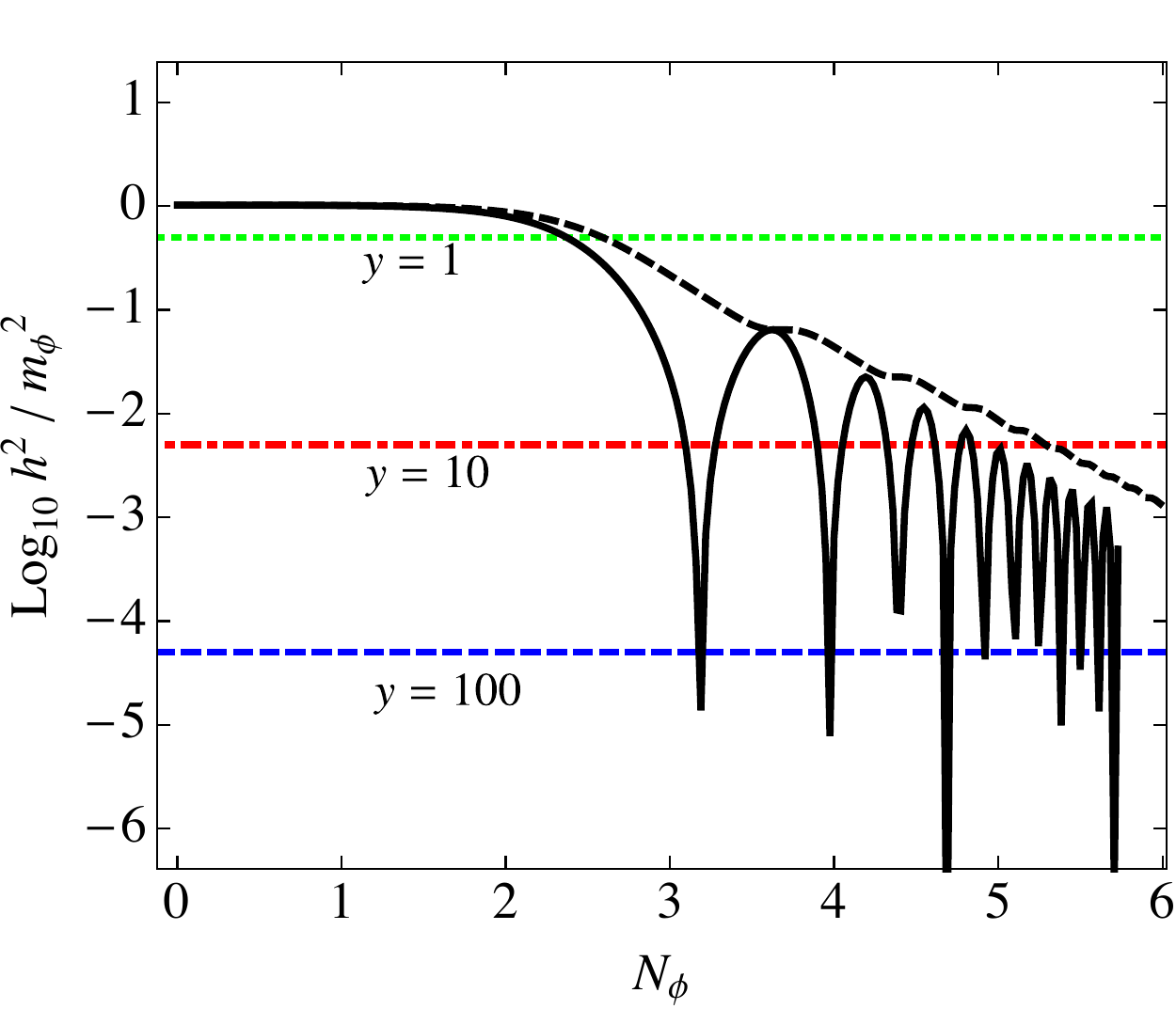}
	\caption{The Higgs field $h^2$ (black solid line) and its mean-squared value $\hrms^2(t)$ (black dashed line) in units of $m_\phi^2$, as a function of the number of $e$-folds $N_\phi$ after the inflaton starts oscillating, obtained as a solution to Eq.~\eqref{eq:Higgs}. Here, we have taken $H_I = m_\phi/2$ and $\lambda_I =10^{-3}$.  We also show the value of $1/2y^2$ for $y = 1$ (green dotted line), $y = 10$ (red dot-dashed line), and $y = 100$ (blue dashed line). Inflaton decay is blocked as long as $4m_f^2 > m_\phi^2$, i.e. when $\frac{h^2}{m_\phi^2} > \frac{1}{2y^2}.$}
	\label{fig:higgs}
\end{center}
\end{figure}

If the Higgs condensate decays too soon after it starts oscillating, then inflaton decay is never blocked, and the effects of this work are not present. Thus we are interested in the case in which the timescale for Higgs decay is longer than the timescale for reheating, i.e., $\Gamma_0\gtrsim H_{\rm dec}$, where $H_{\rm dec}$ is the Hubble scale when the approximate condition for the decay of the Higgs condensate, $m_h^2 \simeq m_{h (W)}^2$ is satisfied.  In the other limit, with $ \Gamma_0 < H_{\rm dec}$, we expect the Higgs condensate to decay away before most of the energy is transferred from the inflaton to the thermal bath, and the blocking effects described here do not significantly alter the reheating process. Note that, while $H_{\rm dec}$ depends on the precise values of $\lambda_I$ and $g$ at the inflationary scale and on the arbitrary direction of the EW symmetry breaking arising from the effective VEV, the decay of the Higgs condensate is largely independent of the perturbative reheating process for $\Gamma_0\gtrsim H_{\rm dec}$. Furthermore, the particular details which determine the exact value of $H_{\rm dec}$ are only relevant when $ \Gamma_0 \simeq H_{\rm dec}$. 
In this subsection we consider only cases in which the inflaton decays completely before the Higgs experiences backreaction from the gauge bosons, i.e. $\Gamma_0 \gg H_{\rm dec}$.  Then Eq.~\eqref{eq:Higgs} suffices to study the effects of effective fermion masses on perturbative reheating. 

Initial conditions for the inflaton are set by the Hubble scale $H_I$ as $\rho_{\phi,I} = 3\mP^2H_I^2/8\pi$, where we use the Planck mass $\mP = 1.221\times 10^{19}\,$GeV. For radiation, initial conditions follow from Eq.~\eqref{eq:radiation_res} as $\rho_{R,I} = 0$, where $\Gamma_\phi = 0$ when the inflaton decay is blocked by particle masses, while $\Gamma_\phi = \Gamma_0$ when the blocking does not occur. Eq.~\eqref{eq:Higgs} determines the dynamics of the time-dependent mass, defined by Eq.~\eqref{eq:fermion_mass}, for the fermion byproducts of perturbative inflaton decay. 

In Fig.~\ref{fig:energydensity}, we show the evolution of $\rho_\phi$ and $\rho_R$ determined by simultaneously solving Eqs.~\eqref{eq:inflaton_res}-\eqref{hubblerate}, where we have taken  $H_I = m_\phi/2$,  $\lambda_I = 10^{-3}$, $\Gamma_0 = 0.1m_\phi$, and $y = 5$.  {While the values of $\Gamma_0$ and $y$ were chosen for clarity in demonstrating the differences between the three treatments of perturbative reheating, similar features arise for all choices of parameters where the effects of Higgs blocking are important and backreaction can be ignored in the evolution in the Higgs field.} The top panel corresponds to  a constant value of $\Gamma_\phi = \Gamma_0$, while the other two panels correspond to the $\Gamma_\phi$ given in Eq.~\eqref{eq:gamma_higgs} for the two limiting cases of $\THi \gg \THu$ (middle panel) and $\THi \ll \THu$ (bottom panel). In all panels, the blue (solid) curve shows $\rho_\phi$, and the red (dashed) curve shows $\rho_R$. In the middle panel, the radiation energy density has a step-function behavior, corresponding to the periods when $h^2 < m_\phi^2 / 2 y^2$ and $\Gamma_\phi \neq 0$, allowing the inflaton field to decay into bursts of SM particles. For the choice of $y=5$ in Fig.~\ref{fig:energydensity} there is only one burst; for higher values of $y$ there would be many such bursts.  Such behavior is not present in the bottom panel, since the decay occurs as the condition in Eq.~\eqref{eq:blocking} is violated and the monotonic $\hrms^2(t)$ falls below the value $1/2y^2$. Defining the reheat time $\td$ through $\rho_\phi(\td) = \rho_R(\td)$, we find $\td \approx 10 m_\phi^{-1}$ for a constant decay rate $\Gamma_0$ (top panel), $\td \approx 70 m_\phi^{-1}$ for the decay rate in Eq.~\eqref{eq:gamma_higgs} and $\THi \gg \THu$ (middle panel), and $\td \approx 450 m_\phi^{-1}$ for the decay rate in Eq.~\eqref{eq:gamma_higgs} and $\THi \ll \THu$ (bottom panel). Inclusion of phase space blocking in Eq. (\ref{eq:gamma_higgs}) delays reheating.

\begin{figure}[h!]
\begin{center}
	\includegraphics[width=\linewidth]{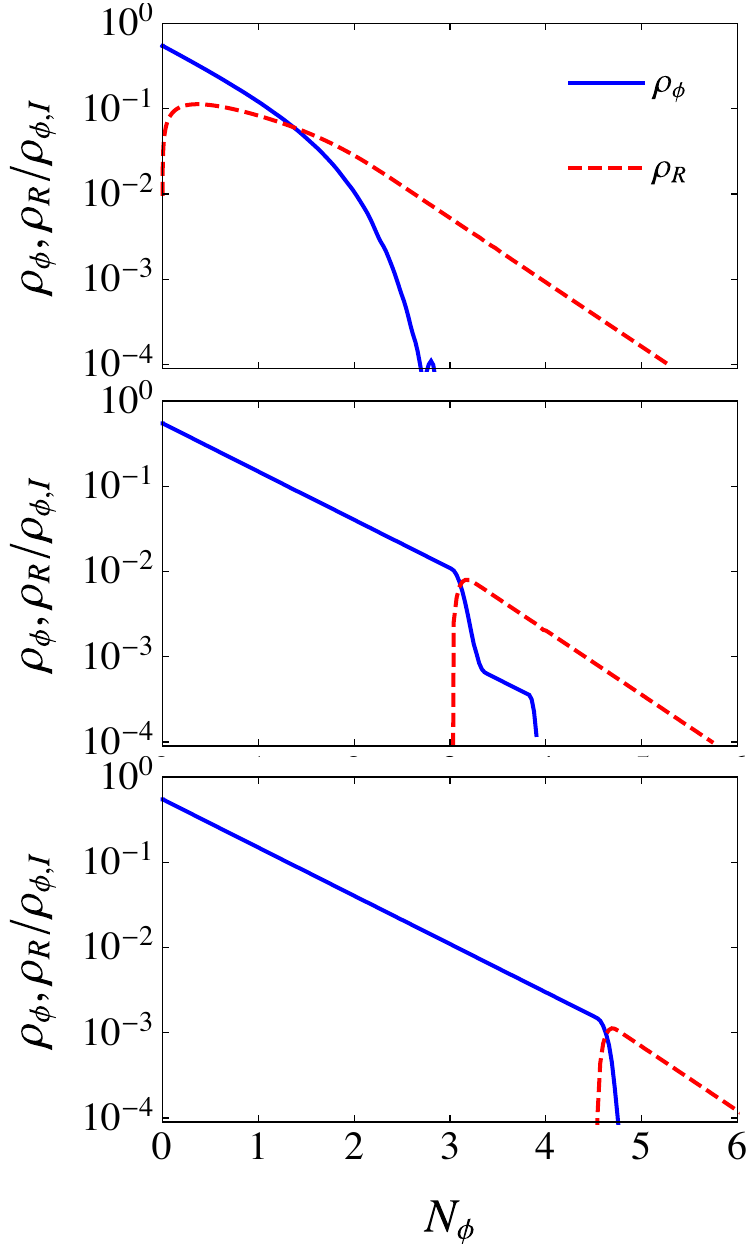}
	\caption{The value of $\rho_\phi$ (blue solid curve) and $\rho_R$ (red dashed curve), in units of $\rho_{\phi,I}$, as a function of the number of $e$-folds $N_\phi$ after the end of inflation, obtained as a solution to the set of Eqs.~\eqref{eq:inflaton_res}-\eqref{hubblerate} for constant $\Gamma_\phi = \Gamma_0$ (top panel) and for $\Gamma_\phi$ given in Eq.~\eqref{eq:gamma_higgs} in the case $\THi \gg \THu$ (middle panel) or $\THi \ll \THu$ (bottom panel). We set $H_I = m_\phi/2$, $\lambda_I = 10^{-3}$, and we take $\Gamma_0 = 0.1m_\phi$ and $y = 5$.  {While the values of $\Gamma_0$ and $y$ were chosen for clarity in demonstrating the differences between the three treatments of perturbative reheating, similar features arise for all choices of parameters where the effects of Higgs blocking are important and backreaction can be ignored in the evolution in the Higgs field.}}	
	\label{fig:energydensity}
\end{center}
\end{figure}

In Fig.~\ref{fig:densityBR}, we show the increase in the number of $e$-folds (relative to the case of no Higgs blocking) $\Delta N_\phi$ by which reheating after inflation is delayed as a function of the parameters $\Gamma_0$ and $y$. The left column shows results obtained by solving the set of Eqs.~\eqref{eq:inflaton_res}-\eqref{hubblerate} (relevant when we do not take  backreaction into account); the right column includes the effects of backreaction onto the Higgs condensate and will be discussed in the next subsection.  Here we have fixed $H_I = m_\phi/2$ and $\lambda_I = 10^{-3}$. Specifically, we show the difference in the number of $e$-folds for the field-dependent $\Gamma_\phi$ in Eq.~\eqref{eq:gamma_higgs} with respect to the results for a constant $\Gamma_\phi$.  The top panel is for the case of  $\THi \gg \THu$ and the bottom panel for $\THi \ll \THu$. Relative to a fixed $H_{\rm dec}$, an earlier inflaton decay leads to a more significant delay in the reheating time. Furthermore, for fixed $m_\phi/ H_I$ and $\lambda_I$, the delay in the reheating time is more significant for a larger $y$ because the time intervals at which the phase space for perturbative decay is open, $h^2 < m_\phi^2 / 2 y^2$, become shorter. The delay in reheating, $\Delta N_\phi$, obtained in the two scenarios typically differ by a factor of order two. For either of the two scenarios, the largest delay in reheating is expected to be $\Delta N_\phi \approx 4.5$ for $y = 10$ and $\Gamma_0 = 0.1m_\phi$, which are the largest values of $y$ and $\Gamma_0 $ we considered.
\begin{figure*}
\includegraphics[width=\linewidth]{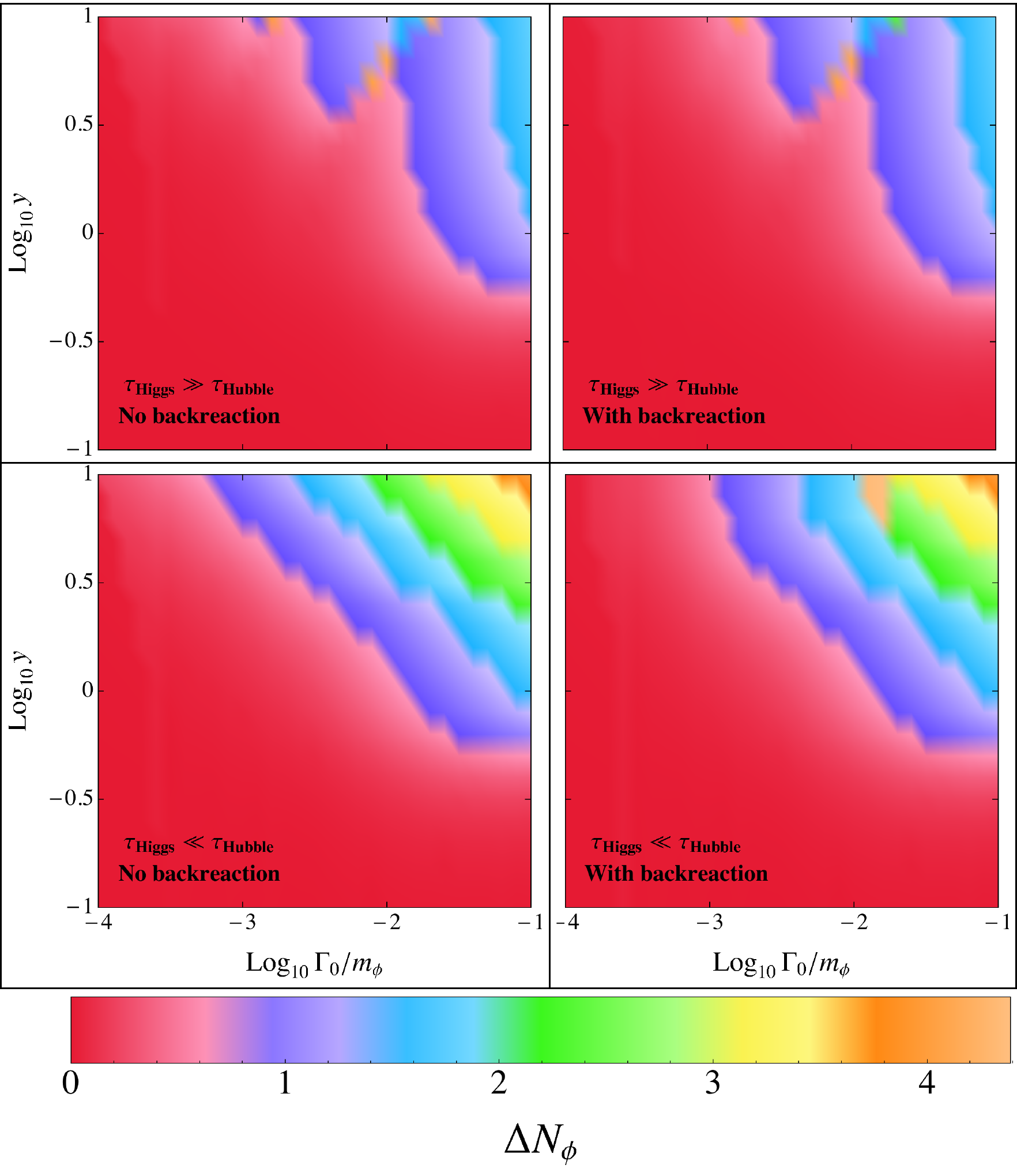}
	\caption{Delay of perturbative reheating due to Higgs blocking, specifically the increase in the number of $e$-folds between the end of inflation and the onset of radiation domination relative to the case without Higgs blocking, as a function of $\Gamma_0$ and $y$.  The two figures in the left column neglect backreaction of the produced gauge bosons on the Higgs field dynamics, 
while the two plots in the right column include backreaction effects.  In the cases with backreaction, the effects of Higgs blocking are less important.  The top two figures are for $\THi \gg \THu$ while the bottom two figures are for $\THi \ll \THu$. Cases with $\THi \ll \THu$ generically have longer delays of reheating due to the effects of Higgs blocking. We see that the strongest effects arise in the case of $\THi \ll \THu$ and are further enhanced for low $\Gamma_0$ if one neglects backreaction (lower left panel).
For  $y = 10$ and $\Gamma_0 = 0.1m_\phi$ the delay peaks at $\Delta N_\phi \approx 4.5$.  {The effects of Higgs blocking on perturbative inflaton decay become less pronounced for smaller values of $y$ and $\Gamma_0$.}}
\label{fig:densityBR}
\end{figure*}

 {
Based on the results presented in Fig.~\ref{fig:densityBR}, one may worry that large Yukawa couplings, which are generally absent in the Standard Model, are needed to yield a significant effect. However, the Yukawa coupling is a proxy for the mass of the fermions, due to the non-zero Higgs VEV, according to Eq.~\eqref{eq:fermion_mass}. If one considers a larger Higgs VEV instead, due to the violations of the de-Sitter equilibrium value for the Higgs as a spectator field (as discussed in Sec.~\ref{sec:HiggsCondensate} and Ref.~\cite{Hardwick:2017fjo}), a smaller Yukawa coupling is sufficient to produce a similar result. The effect of a larger value of the Higgs VEV can be visualized by shifting  the black curves of Fig.~\ref{fig:higgs} upwards. In that case, a larger delay in the reheating time would arise for each value of the Yukawa coupling. In concslusion, the Yukawa values shown in Fig.~\ref{fig:densityBR} must be accompanied by the assumption of a de-Sitter equilibrium solution for the Higgs field until the end of inflation. Otherwise, they have to be properly rescaled, by the ratio of the higgs VEV to the Hubble scale.}

\subsection{Including backreaction}

So far, we have neglected the backreaction of the production of the gauge bosons on the Higgs field dynamics. This approximation is expressed in Eq.~\eqref{eq:Higgs}, where the excitation of the gauge bosons does not affect the evolution of the Higgs field. Depending on the couplings $g$ and $\lambda_I$,
the inclusion of backreaction could change the conclusions drawn so far. In this section we will choose value of $g$ high enough to demonstrate the effect of backreaction, but note that for a SM value evaluated at inflationary scales backreaction effects may be insignificant. 

We sketch the effect due to the inclusion of gauge boson production by introducing the effective mass of the gauge boson $m_{h (W)}^2$, as defined in Eq.~\eqref{eq:mass_Wboson}. When $m_{h (W)}^2$ equals the effective mass of the Higgs field, $m_h^2 = 3\lambda_I h^2$, we assume that the Higgs field decays and the decay rate of the inflaton is no longer blocked, reaching the value $\Gamma_0$. For this, we consider a set of Boltzmann equation analogous to what we have previously discussed in Eqs.~\eqref{eq:inflaton_res}-\eqref{eq:radiation_res},
\ba
	\dot{\rho}_\phi +3H\rho_\phi &=& -\Gamma_{\rm BR} \rho_\phi, \label{eq:inflaton_res_br}\\
	\dot{\rho}_R + 4H\rho_R &=& \Gamma_{\rm BR} \rho_\phi,\label{eq:radiation_res_br},
\ea
where the decay rate $\Gamma_{\rm BR}$ is expressed in terms of $\Gamma_\phi$, see Eq.~\eqref{eq:gamma_higgs}, and $\Gamma_0$ as
\be
	\Gamma_{\rm BR} \!=\! \Gamma_\phi\Theta\!\(3\lambda_I h^2 \!-\! m_{h (W)}^2\!\) + \Gamma_0\Theta\!\(m_{h (W)}^2 \!-\! 3\lambda_I h^2\!\).
\ee
The expression for $\Gamma_{\rm BR}$ above states that the Higgs blocking only affects the dynamics during the time period when backreaction can be neglected, $m_{h (W)}^2 \lesssim 3\lambda_I h^2$. We solve Eq.~\eqref{eq:gauge} in the illustrative case $g \simeq 2.06$, i.e., $q_W \equiv g^2/4\lambda_I \simeq 1060$, where $q_W$ is  the resonance parameter which arises from writing the equation of motion of the W-boson as a Mathieu equation, see for example Ref.~\cite{Greene:1997fu}. 
Although the value of the gauge coupling $g$ is somewhat higher than what one would expect in the SM by a factor 4-5, we choose such a high value in order to illustrate the effects of backreaction on the Higgs blocking, in the case of the perturbative inflaton decay to fermions. However, even for a smaller value of $g \simeq 0.5$, the effects of backreaction could still be present for values of $y$ which are outside of the perturbative regime\footnote{As shown in the lower panels of Fig.~\ref{fig:densityBR}, backreaction only becomes important for $ y \gtrsim 1$. This is because, as shown in Fig.~\ref{fig:higgs}, larger induced fermion masses will lead to more efficient phase space blocking and larger $\Delta N_\phi$. Thus, only for sufficiently large $y$ is the blocking in effect long enough for the decay time of the inflaton to become similar the decay time of the Higgs. If we had calculated the decay of the Higgs with $g \simeq 0.5$, then the Higgs would have decayed later, requiring an inflaton Yukawa coupling larger than allowed by perturbative unitarity, $y \gtrsim 4\pi$, in order for the effects of backreaction to be important.}. In principle, the effective gauge boson mass is obtained by computing the expression in Eq.~\eqref{eq:mass_Wboson}, which involves an integration over the gauge boson momenta $k$. Here, we simplify such computation by considering that, for $q_W \gg 1$, all modes $k$ are excited up to the cutoff scale~\cite{Greene:1997fu}
\be
	k_* \equiv \sqrt{\lambda_I} h_{\rm osc}\,a_{\rm osc} \,\(\frac{q_W}{2\pi^2}\)^{1/4}.
	\label{eq:cutoffk}
\ee
This approximation is valid when the solutions to Eq.~\eqref{eq:gauge} lie in a resonant band, as it is for our choice of $g$, and allows us to write the occupation number as
\be
	n_k = n_0\,\Theta\(k-k_*\),
	\label{eq:approx_nk}
\ee
where $n_0$ is the occupation number when assuming $k = 0$. Using this expression for $n_k$ to evaluate the
 integral in Eq.~\eqref{eq:mass_Wboson} , we
 obtain the induced mass of the W boson as
\be
	m_{h (W)}^2 \approx \frac{4\pi k_*^3}{3} \frac{\lambda_I q_W n_0}{(2\pi a)^3\sqrt{q_W \lambda_I} h} = \frac{\sqrt{2}\lambda_I^2 n_0}{6\pi} \left(\frac{ q_W }{2\pi^2}\right)^{5/4} h^2,
	\label{eq:mass_Wbosons1}
\ee
where in the last expression we have used Eq.~\eqref{eq:cutoffk} and the relation $h\propto a^{-1}$, which is valid after the Higgs has started oscillating.  We have checked that the result in Eq.~\eqref{eq:mass_Wbosons1}, obtained in the $k = 0$ approximation, is consistent with the numerical integration in Eq.~\eqref{eq:mass_Wboson}. Backreaction becomes important once $m_{h (W)}^2 \simeq 3\lambda_I h^2$,  or when the occupation number
 is\footnote{This result differs from Eq.~(4.10) in Ref.~\cite{Enqvist:2013kaa}, where the authors obtain a time-dependent occupation number. 
This difference is related to the inconsistency pointed out in Footnote~\ref{footnote_Enqvist}.} 
\be
	n_0 = \frac{9\pi\sqrt{2}}{\lambda_I}\left(\frac{2\pi^2}{ q_W }\right)^{5/4}.
	\label{eq:findbackreaction}
\ee
We use Eq.~\eqref{eq:numberdensityW} to obtain an additional expression for $n_0$. For this, we use the $W$ field obtained by solving Eq.~\eqref{eq:gauge} with $k=0$. We find that $n_0(t)$ can be approximated by~\cite{Enqvist:2013kaa}
\be
	n_0(t) = \exp\(\mu_0 \sqrt{m_\phi (t-t_{\rm osc})}\),
	\label{eq:approxW}
\ee
where $\mu_0 = 0.185$. We match the expressions in Eqs.~\eqref{eq:findbackreaction} and~\eqref{eq:approxW} to obtain the time at which backreaction becomes important, or
\be
	t_{\rm dec} = t_{\rm osc} + \frac{1}{m_\phi}\(\frac{\ln n_0}{\mu_0}\)^2 \approx t_{\rm osc} + 270/m_\phi,
	\label{eq:time_backreaction}
\ee
corresponding to $N_\phi \approx 5$ $e$-folds after the inflaton field has started to oscillate. Including backreaction in the Higgs dynamics modifies the results obtained in the previous section, which can be seen when comparing the left panels of Fig.~\ref{fig:densityBR} to the right panels in Fig.~\ref{fig:densityBR}, where there is a cutoff introduced in the delay of reheating, $\Delta N_\phi$. Backreaction affects the region $y \gtrsim 1$, where the delay of reheating becomes less pronounced.

\subsection{Including gauge boson decays to fermions}
\label{sec:bosonstofermions}

The mass of the gauge bosons produced when the inflaton field stars to oscillate is proportional to the value of the Higgs field at that time, as shown in Eq.~\eqref{eq:motion_Wboson}. In the language of Ref.~\cite{felder1999}, the gauge bosons ``fatten'' as the value of the Higgs field increases. If the gauge bosons couple to fermions, they would quickly dissipate into these lighter degrees of freedom. This mechanism has been discussed in Refs.~\cite{bezrukov2009, garciabellido2009} in the context of Higgs inflation within the Standard Model, while a full lattice computation has been recently deployed in Ref.~\cite{repond2016}. This scenario in a supersymmetric version of the Standard Model has been discussed in Ref.~\cite{allahverdi2011}. 

 {In the model we consider, the inflaton field is not the Higgs field; nevertheless, the gauge bosons will couple to fermions, with an additional interaction term in the Lagrangian 
arising from the covariant derivative in the kinetic term for $\psi$,
\be
	\LL_{\rm int} = -i g \,W_\mu\,\bar\psi \gamma^\mu \psi,
\ee
where $g$ is the $SU(2)$ coupling constant. This interaction leads to the decay rate
\be
	\Gamma_{W \to \bar\psi \psi} = \frac{g^2\,m_W}{48\pi} = \frac{g^3\,h}{96\pi},
\ee
where we have set $m_W = g h/2$ and assumed the fermions are effectively massless. As discussed in Ref.~\cite{Enqvist:2014tta}, the decay of gauge bosons to fermions can suppress the resonant decay of the Higgs condensate if 
\be
{1 \over \Gamma_{W \to \bar\psi \psi}} \lesssim \THi \sim {1 \over m_{\rm osc}} ,
\ee
which would allow the gauge bosons to decay in between oscillations of the Higgs field. Since $h_{\rm osc} \sim m_{\rm osc}$ and we assume $g \simeq 2$, the gauge bosons can indeed perturbatively decay to fermions on time scales similar to that of Higgs oscillations.}

{In the limiting cases we consider, one would therefore expect the decay of the Higgs condensate to be delayed when $\THi \gg \THu$ while the effects of backreaction should remain essentially unchanged when $\THi \ll \THu$. More specifically, the effects of Higgs blocking on perturbative reheating when $\THi \gg \THu$ and backreaction to the Higgs condensate decay is considered should more closely resemble the corresponding case when backreaction is ignored. Thus, as with the choice of $g \simeq 2$, we are making a conservative estimate of Higgs blocking effects in the case of $\THi \gg \THu$ when backreaction is included.}

\subsection{Effects on the reheat temperature}
\label{sec:reheatTemp}

In previous subsections, we have calculated the delay in reheating in terms of $\Delta N_\phi$ independently of the precise value of $m_\phi$. However, in a realistic model of inflation we want to achieve the correct density perturbations and translate delays in reheating into predictions for the reheat temperature. To be illustrative, we will consider a quadratic shape for the form of the inflaton potential near its minimum and an inflaton mass $m_\phi \sim 10^{13}\,$GeV.  A quadratic potential with this mass throughout inflation for all values of $\phi$ would produce the correct  density perturbations; yet it is ruled out since it yields a value of $r\sim 0.2$, placing it outside the current experimental bounds. However, it is possible to construct models where the inflaton potential is non-quadratic during the period when density fluctuations and tensor modes are produced (60-50 $e$-folds before the end of inflation), and yet is quadratic near the bottom (e.g. some variants of axion monodromy~\cite{McAllister:2008hb}). Henceforth to be concrete we will assume a quadratic potential near the minimum and mass $m_\phi \sim 10^{13}\,$GeV.


When the inflaton field decays, the Universe transitions to a radiation dominated state with the Hubble rate
\be
	H(T) = \sqrt{\frac{8\pi \,\rho_R}{3\mP^2}} = \sqrt{\frac{4\pi^3}{45}g_*(T)}\frac{T^2}{\mP},
	\label{eq:hubblestandard}
\ee
where $g_*(T)$ is the number of the relativistic degrees of freedom at temperature $T$. We compute the reheating temperature $\TRH$ by first evaluating the Hubble rate $H(\TRH)$ when $\rho_\phi = \rho_R$ and then using Eq.~\eqref{eq:hubblestandard}. Clearly, a prolonged inflaton oscillation (matter domination) stage affects the reheating temperature $\TRH$. In general, we expect that $\TRH$ lowers when the Higgs blocking effect is present, because the coherent inflaton oscillation period lasts longer for the same value of $H_I$.

So far, we have obtained the results by considering the initial value of the Higgs field $h_I$ to be the ``central value'' in Eq.~\eqref{eq:centralvalue}. However, for a given value of $H_I$, the initial value of the Higgs field is distributed according to the PDF in Eq.~\eqref{eq:pdfhiggs}, so in some Hubble patches the value of $h_I$ can differ greatly from what is expected by its central value. For each value of $h_I$, we obtain a different reheat temperature $\TRH$, hence the reheat temperature of each Hubble patch of our observable Universe will depend on the PDF $f_{\rm eq}(h)$. 
In particular, the probability of finding the reheat temperature in a Hubble patch between $T_1$ and $T_2$ is
\be
\label{eq:TrhCDF}
P(T_1<\TRH<T_2) = \int_{T_1}^{T_2} \tilde f(\TRH) d\TRH ,
\ee
where the PDF for the reheat temperature is given by
\be
	\tilde f(\TRH) \equiv f_{\rm eq}(h_I(\TRH)) \left |{dh_I\over d\TRH} \right | .
	\label{eq:TrhPDF}
\ee
We solve the set of the Boltzmann Eqs.~\eqref{eq:inflaton_res}-\eqref{hubblerate} for given $H_I = m_\phi/2$ and $\Gamma_0$, obtaining the one-to-one relation $h_I \equiv  h_I(\TRH)$ without the effects of backreaction included. We also solve Eqs.~\eqref{eq:inflaton_res_br}-\eqref{eq:radiation_res_br} and Eq.~\eqref{hubblerate} when backreaction is included. Since we also consider the evolution of the Higgs with either $\THi \gg \THu$ or $\THi \ll \THu$, we obtain a PDF of the reheat temperature for each of the four cases considered before. Since, for cases with $\THi \gg \THu$, the numerical function $h_I(\TRH)$ has oscillatory features, we have fit to a polynomial in order to obtain a smooth derivative for the calculation of $\tilde f(\TRH)$ in Eq.~\eqref{eq:TrhPDF}. Although we expect the oscillations of $h_I(\TRH)$ to manifest as features in $\tilde f(\TRH)$, we note that these features would be less apparent in a more realistic treatment of the Higgs field, with $\THi \simeq \THu$, and leave a calculation of the associated effects on $\tilde f(\TRH)$ for future work.  

We plot results in Fig.~\ref{fig:plotPDF}, where we show the PDF $\tilde f(\TRH)$ defined in Eq.~\eqref{eq:TrhPDF} for $\THi \gg \THu$ (top row panels) and $\THi \ll \THu$ (bottom row panels), as well as when backreaction is neglected (left column panels) and when its effect is included (right column panels). For each panel, the different lines represent different values of $\Gamma_0 = 10^{-3} m_\phi$ (red solid line), $\Gamma_0 = 10^{-2} m_\phi$ (blue dotted line), and $\Gamma_0 = 10^{-1} m_\phi$ (green dot-dashed line). For comparison, the reheat temperature without any blocking effects $\TRH^0$ is obtained from the Hubble equation, given by Eq.~\eqref{eq:hubblestandard} as $3 H(\TRH^0) = \Gamma_0$, or
\begin{align}
	\TRH^0  = \left(\frac{5}{4\pi^3g_*}\right)^{1/4}\!\sqrt{\Gamma_0\,\mP} \approx 0.14 \left ( {100 \over g_*} \right )^{1/4} \sqrt{\Gamma_0 \mP} .
	\label{eq:Trh0}
\end{align}
Setting $g_* = 106.75$ and $m_\phi = 10^{13}\,$GeV, we find $\TRH^0 = 4.9 \times 10^{13} \,$GeV when $\Gamma_0 = 10^{-3}m_\phi$, $\TRH^0 = 15 \times 10^{13} \,$GeV for $\Gamma_0 = 10^{-2}m_\phi$, and $\TRH^0 = 49 \times 10^{13} \,$GeV for $\Gamma_0 = 10^{-1}m_\phi$. 

As stated previously, to be concrete, we have taken a quadratic potential near the minimum and mass $m_\phi \sim 10^{13}\,$GeV.  We note that, for values of the inflaton mass within several orders of magnitude of this number and a fixed ratio of $ m_\phi / H_I$, our results for the relative delay in reheating remain roughly the same:  the ratio of $\TRH / \TRH^0$ changes by factors up to $\sim 2$ (compared to the results in Fig.~\ref{fig:plotPDF}).

The results in Fig.~\ref{fig:plotPDF} are consistent with the features of Fig.~\ref{fig:densityBR}. Naively, for a fixed delay in the reheating, an increase in the value of $\Gamma_0$ leads to higher $\TRH$. In contrast, as seen in  Fig.~\ref{fig:densityBR}, $\Delta N_\phi$ increases with higher $\Gamma_0$ for a fixed value of $y$. Both these effects are present when considering the respective PDFs for different $\Gamma_0$ in the top-left panel of Fig.~\ref{fig:plotPDF}, when $\THi \gg \THu$ and without considering the effects of backreaction. When increasing $\Gamma_0$, the most likely values of $\TRH$ are higher. At the same time, the overall delay in reheating from $\TRH^0$ (as shown in the legend of Fig.~\ref{fig:plotPDF}) to the most likely values of $\TRH$ also increases.

From the the bottom-left panel of Fig.~\ref{fig:plotPDF} (no backreaction but $\THi \ll \THu$), we again see the scenario where $\THi \ll \THu$ generically leads to an even more delayed reheating, as the most likely values of $\TRH$ decrease for all $\Gamma_0$ relative to the previous case with $\THi \gg \THu$. Also, the largest decreases in the most likely values of $\TRH$, relative to the case with $\THi \gg \THu$, occur for larger values of $\Gamma_0$.

As shown in the right panels of Fig.~\ref{fig:plotPDF}, backreaction inhibits the blocking effect and restores the standard unblocked reheating picture, thus it generally predicts higher reheat temperatures. However, since backreaction only acts when $\Gamma_0 \lesssim 0.01m_\phi$ with our choice of parameters, the PDF for the red solid curve (corresponding to $\Gamma_0 = 10^{-3}m_\phi$) in the top-right panel predicts that higher reheat temperatures are more likely than in the top-left panel. The blue and green lines remain identical because backreaction does not effect reheating with these choices of $\Gamma_0$.
\begin{figure*}
	\includegraphics[width=\linewidth]{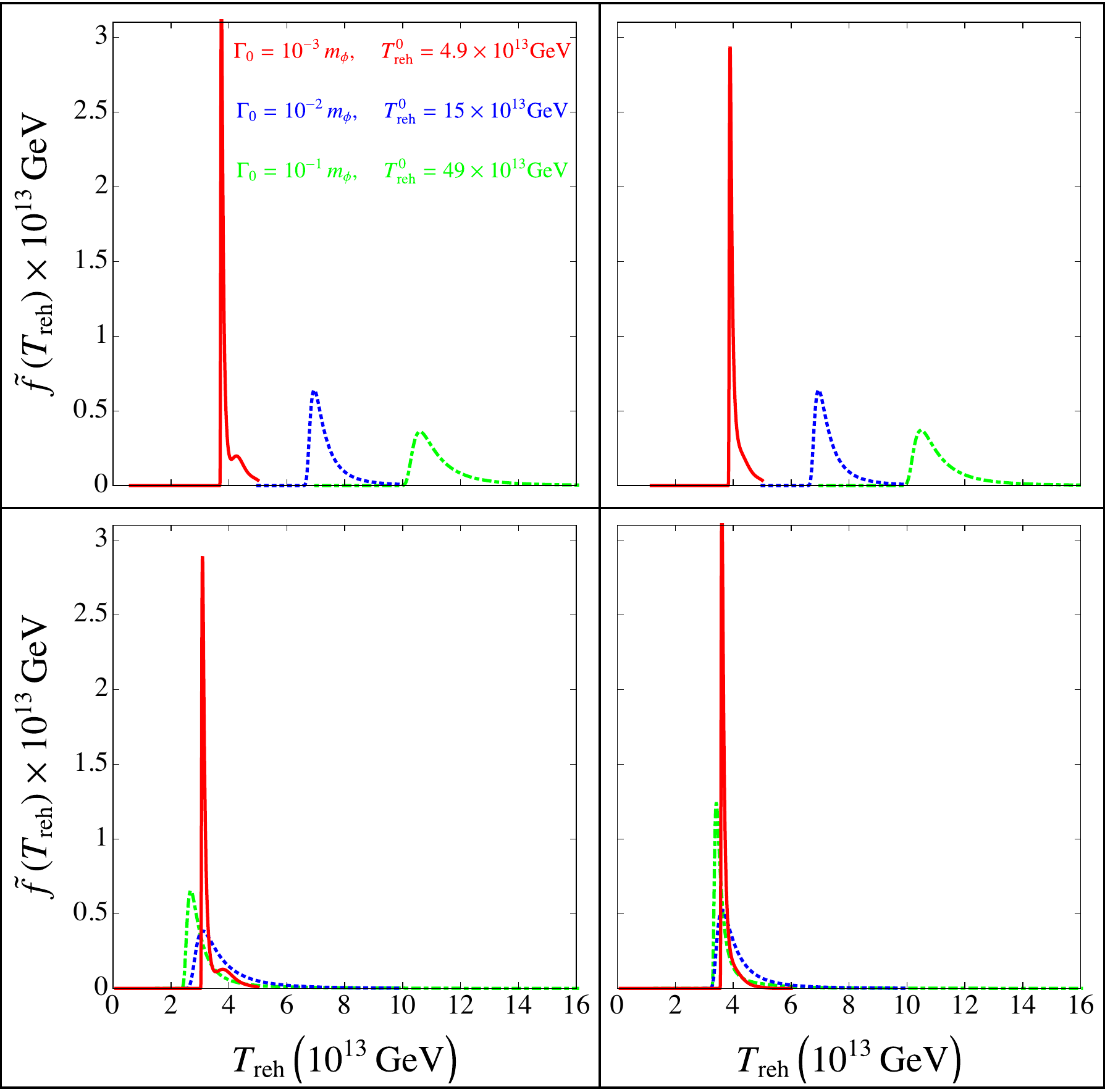}
	\caption{Lowered reheat temperature due to Higgs blocking for the case of perturbative inflaton decay, assuming $m_\phi = 10^{13}\,$GeV.  The PDF $\tilde f(\TRH)$ defined in Eq.~\eqref{eq:TrhPDF} is shown as a function of the reheating temperature $\TRH$ for either $\THi \gg \THu$ (top row) or $\THi \ll \THu$ (bottom row) and when backreaction is either neglected (left column) or considered (right column). The values of $\TRH^0$ in the legend are computed in the case when the blocking effect is absent. We consider different values for $\Gamma_0 = 10^{-3} m_\phi$ (red solid line), $\Gamma_0 = 10^{-2} m_\phi$ (blue dotted line), and $\Gamma_0 = 10^{-1} m_\phi$ (green dot-dashed line). One can see that the Higgs blocking considered in this paper can lead to a reheat temperature that is suppressed by roughly an order of magnitude compared to the standard case without blocking (the curves in the figures can peak at a lower values than the unblocked numbers in the legend). We note that, for values of the inflaton mass within several orders of magnitude of $m_\phi = 10^{13}\,$GeV and a fixed ratio of $ m_\phi / H_I$, our
results for the relative delay in reheating remain roughly the same:  the ratio of ${\TRH / \TRH^0}$ changes by
factors up to $\sim 2$ (compared to the results in this figure).}
	\label{fig:plotPDF}
\end{figure*}

The main result of this section is the following:  In the case of perturbative inflaton decay, the Higgs blocking considered in this paper can lead to a reheat temperature that is suppressed by roughly an order of magnitude compared to the standard case without blocking (the curves in Fig.~\ref{fig:plotPDF}  can peak at a lower values than the unblocked numbers in the legend).

\section{Gauge Preheating} \label{sec:gauge}

\subsection{Resonant gauge boson production}

If, instead of perturbative reheating through a Yukawa coupling of the inflaton to SM fermions, we consider Chern-Simons couplings of the inflaton to SM gauge bosons, resonant gauge boson production can reheat the Universe almost instantaneously after the end of inflation for sufficiently large values of the inflaton-gauge couplings~\cite{Adshead:2015pva, Adshead:2016iae}. Although previous analyses of gauge preheating have considered effectively massless gauge bosons, we analyze the effects of a large gauge boson mass on the structure of the resonance.
 We assume an axion-like derivative coupling of the inflaton to an Abelian gauge boson with an effective mass given by the coupling of the gauge boson to the SM Higgs\footnote{Breaking of the Electroweak symmetry during inflation would leave one of the resulting gauge bosons massless. In a general effective field theory setting, the inflaton would couple to both the massive and the massless gauge bosons alike. If one considers a definite UV completion of the SM, the couplings of the inflaton to the separate gauge bosons would depend on the charges of the fermions under the SM and the Peccei-Quinn-like symmetry which provides for the shift symmetry of the axion~\cite{Freese:1990rb, Adams:1990pn}. Such an analysis is beyond the scope of the current paper. We instead consider only an effectively massive $U(1)$ gauge field, as an indicative case of a more complicated and model-dependent process.}.
Chern-Simons couplings to gauge fields are generic for example in models of a pseudo scalar inflaton as in Natural Inflation~\cite{Freese:1990rb, Adams:1992bn}, since they respect the underlying shift symmetry. 
The relevant terms in the potential of the corresponding effective Lagrangian are given by
\ba
	V_A = {\alpha \over 4 f} \phi F^{\mu \nu} \tilde{F}_{\mu \nu} + {M^2 \over 2} A^\mu A_\mu ,
	\label{eq:V_A}
\ea
where $F_{\mu \nu} = \partial_\mu A_\nu - \partial_\nu A_\mu$ and $\tilde{F}_{\mu \nu} = \epsilon_{\mu \nu \beta \gamma} F^{\beta \gamma}$. 
Note that the electroweak symmetry breaking due to the large Higgs condensate, which develops when the Higgs is a spectator field during inflation, yields the mass term for the gauge boson $M^2 \sim g '^2 h^2$, where $g'$ is the Abelian gauge coupling\footnote{A detailed derivation of the equations of motion is given in Appendix \ref{sec:abelianhiggsedmodel}.}. While the dynamics of the Higgs field are still technically determined by Eq.~\eqref{eq:Higgs}, the Higgs field does not begin oscillating until well after gauge preheating, which only lasts for a few $e$-folds after the end of inflation. We thus take the gauge boson mass to be constant during the last few $e$-folds of inflation and immediately thereafter and depend only on the Hubble scale at the end of inflation.

In general, a Chern-Simons coupling to gauge bosons can induce a coupling of the inflaton to the axial vector current,
\begin{align}
	{\cal L}_{\rm \phi\bar\psi\psi} \sim i {1\over f} \partial_\mu\phi \bar\psi \gamma_5\gamma^\mu\psi \, .
\end{align}
Even though such couplings can source interesting phenomenology~\cite{Adshead:2015kza, Adshead:2015jza}, Pauli's exclusion principle does not allow for sufficient transfer of energy from the inflaton to fermions during preheating. We will thus neglect these processes here.

The equation of motion for the gauge boson products of the inflaton decay is similar to that of the gauge boson products of the Higgs condensate decay in Eq.~\eqref{eq:gauge} and is given by~\cite{Adshead:2015pva}
\ba
	\ddot \chi^\pm_k  + \omega^2 \chi^\pm_k= 0, ~ \omega^2 =  {k^2 \over a^2}  \mp { \alpha\over f} {k\over a} \dot \phi {+ M^2 }  { +  { \dot a^2 \over 4a^2}    - {\ddot a\over 2a} } ,
	\label{eq:pregauge}
\ea
where $A_k$ is the Fourier transform of the transverse component of the gauge field and $\chi_k = a^{1/2}A_k$. Note that, since the last two terms in the definition of $\omega^2$ are subdominant after inflation, we will neglect them in all analytic estimates although we still include them in the numerical calculations. The mass term $M^2 \sim g '^2 h^2$ is equivalent to the second term $g^2 h^2/4$ on the right hand side of Eq. (\ref{eq:motion_Wboson}). Using Eq. (\ref{eq:centralvalue}), we may write $M/H_I \sim g' \lambda_I^{-1/4} $ so that we may use the quantity $M/H_I$ as an effective coupling constant. We can use the WKB approximation to solve Eq.~\eqref{eq:pregauge} in regions where the effective frequency is slowly varying, $\dot \omega / \omega^2 \ll 1$.   

Preheating begins with a period of tachyonic instability, with $\omega^2(t) <0$. At the end of this period, the coefficient of the solution which describes the amplitude of modes is given by~\cite{Adshead:2015pva}
\ba
	\chi_k  ={1\over \sqrt{2k}}e^{X_k}  ~  , ~ X_k = \int_{t_1}^{t_2} \Omega_k(t') dt' ~,~ \Omega_k^2 = -\omega^2 ,
	\label{eq:WKB}
\ea 
where the time period during which $\omega^2(t)<0$ is taken to be $t_1<t<t_2$\footnote{If one roughly thinks of Eq. (\ref{eq:pregauge}) as a harmonic oscillator equation, one can think of $\omega^2<0$ as the regime of exponential growth rather than oscillation. Since it is not exactly an harmonic oscillator, we use the terminology ``effective frequency."}.
Subsequent to the first tachyonic burst, the effective frequency acquires an oscillatory component, due to the fact that the inflaton condensate is oscillating around the minimum of its potential. This can  lead to a period of parametric resonance and further particle production for later times  $t > t_2$. For the massless gauge boson case, the parametric resonance was discussed for example in Ref.~\cite{Adshead:2015pva}.


In Ref.~\cite{Adshead:2016iae} it was shown that gauge preheating dynamics can be divided into three cases, depending on the size of the axion-gauge coupling ${\alpha / f}$ and the resulting effects:\footnote{The values of the couplings mentioned in this section refer to a model with a quadratic inflaton potential. For shallower potentials, like axion monodromy, the values of the couplings in each regime are increased, but the characteristics of these regions persist. A numerical comparison between the cases of quadratic and axion monodromy potential was performed in Ref.~\cite{Adshead:2015pva}.} 

\begin{enumerate}
\item Small coupling $\left (\alpha/f\lesssim 9\,\mP^{-1} \right )$: The analysis can be done entirely using linear equations of motion, neglecting backreaction  effects and mode-mode coupling. The resulting fraction of the energy density that is transferred non-perturbatively into the gauge fields is a few percent or less.
\item Intermediate coupling $\left ( 9\,\mP^{-1} \lesssim \alpha/f\lesssim  10\, \mP^{-1} \right )$: The linear equations of motion can be used and provide very accurate results, until the time when the energy density of the gauge fields becomes comparable to the energy density of the background inflaton field. In this case preheating can be complete and lasts typically a few inflaton oscillations and a few $e$-folds.
\item Large coupling $\left (\alpha/f\gtrsim 10 \,\mP^{-1} \right )$: In this case the entirety of the energy density of the inflaton is transferred onto the gauge fields within one axion oscillation, driven by an initial period of tachyonic amplification, henceforth denoted as the first tachyonic burst.
 Proper study of this case requires the use of lattice simulations, which is beyond the scope of the present work.
\end{enumerate} 

We begin by closely examining the case of ${\alpha / f} = 9\mP^{-1}$, which was considered as the typical example of the ``intermediate coupling case" in Ref.~\cite{Adshead:2016iae}. We use a linear no-backreaction analysis, an approach that has previously been shown to agree very well with the full lattice results during the initial stages of preheating.  Indeed, we show that this analysis also works equally well for the subsequent evolution when we introduce a non-zero mass due to the Higgs mechanism.

Fig.~\ref{fig:omega2} shows the square of the effective frequency $\omega^2$ as a function of the number of $e$-folds relative to the end of inflation (the end of inflation is here taken to be at $N_\phi = 0$), for ${\alpha / f} = 9\mP^{-1}$. We show results for different values of the ratio $M/ H_I$ ranging from zero to one: as described before, this ratio may be thought of as an effective coupling constant. The upper panel shows the first tachyonic burst, which takes place during the last few $e$-folds of inflation; the lower panel shows the parametric resonance right after inflation ends (right after $\omega^2=0$ for the first time). In the upper panel, the behavior of $\omega^2$ during the first tachyonic burst is qualitatively independent of the value of $M/ H_I$: the negative effective frequency-squared differs by only about $10\%$ for typical values of the gauge boson mass. However, $\omega^2$ behaves differently for different values of the ratio $M/ H_{I}$ during the parametric resonance after the first inflaton zero-crossing, as shown in the lower panel of Fig.~\ref{fig:omega2}. In fact, for larger values of this ratio, the oscillation of the effective mass-squared acquires a constant positive offset, leading to a severe suppression of the parametric resonance.  


The behavior just described can be explained in the language of the Mathieu equation $\ddot \chi^\pm_k + \omega^2 \chi^\pm_k =0$, where we consider the effective frequency given by Eq.~\eqref{eq:pregauge} and neglecting the last two terms arising from the expansion,
\ba
	{\omega^2\over H^2} =\left( {k\over aH} \right)^2 \mp {\alpha\over f} {k \over aH}{ \dot \phi \over H} + \left ( {M\over H}\right)^2 \, .
	\label{eq:omegaoverH}
\ea
The expression for the effective frequency in Eq.~\eqref{eq:omegaoverH} allows for a very simple estimation of the effect of the Higgs mass during the initial tachyonic amplification phase. Considering the mode that exits the horizon at the end of inflation $k=aH$, and using a quadratic inflaton potential for the illustrative purpose, we obtain $\dot \phi = m_\phi \phi/\sqrt{2} \simeq m_\phi \mP /\sqrt{2} $ and $H_I \simeq m_\phi/2$ at the end of inflation. Inserting these expressions into Eq.~\eqref{eq:omegaoverH} and using the condition ${\alpha / f} = 9\mP^{-1}$, which is in the intermediate coupling regime, we get the simple estimate
\ba
	{\omega^2\over H_I^2}  \simeq -5.4 + \left ({M\over H_I} \right )^2  \, ,
	\label{eq:omegaoverH1}
\ea
where generically $M/ H_I = {\cal O}(1)$. This is a useful estimate of the effect of the gauge field mass on the tachyonic amplification. However, we must also consider the point where the effective frequency-squared acquires its largest (negative) value. By using the slow-roll expression for $\dot \phi$, it is easy to compute
\ba
{  \omega_{\rm min}^2\over H_I^2} \simeq -4\left (  {\alpha \over f} {\mP\over \sqrt{6}} \right ) + \left ({M\over H_I} \right )^2
\label{eq:omegaoverH2}
\ea
Setting ${\alpha / f} = 9\mP^{-1}$, we see that the effect of the gauge field mass at this point is less than $10\%$ for $M\lesssim H_I$ and rises to over $25\%$ for $M=2 H_I$. Hence it is reasonable to expect that massive gauge fields exhibit similar early tachyonic behavior to their massless counterparts, especially for masses that do not significantly exceed the Hubble scale. The estimations given in Eq.~\eqref{eq:omegaoverH1} and \eqref{eq:omegaoverH2} agree with the plot shown in Fig. \ref{fig:omega2}.

\begin{figure}[h!]
\begin{center}
	\includegraphics[width=\linewidth]{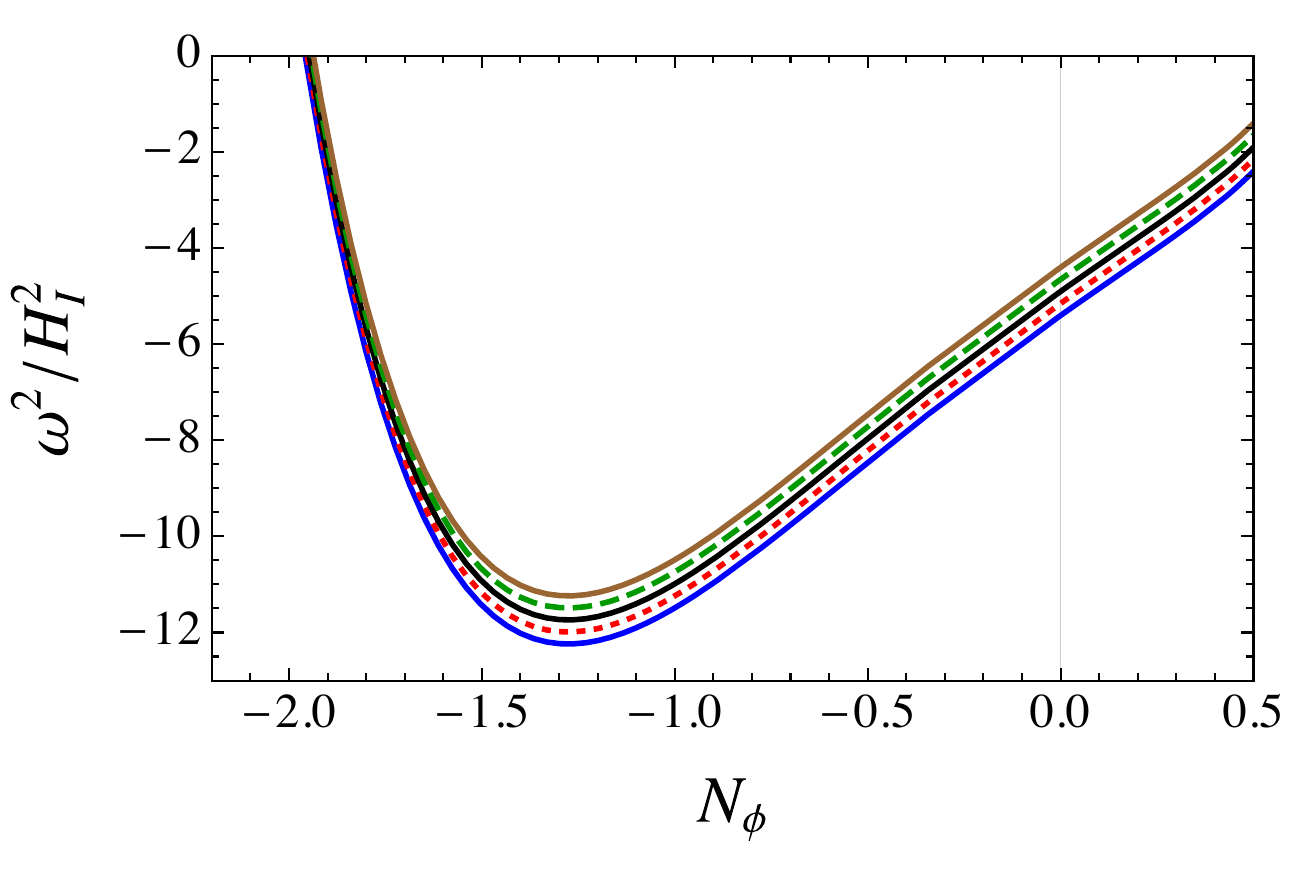}
	\\
	\includegraphics[width=\linewidth]{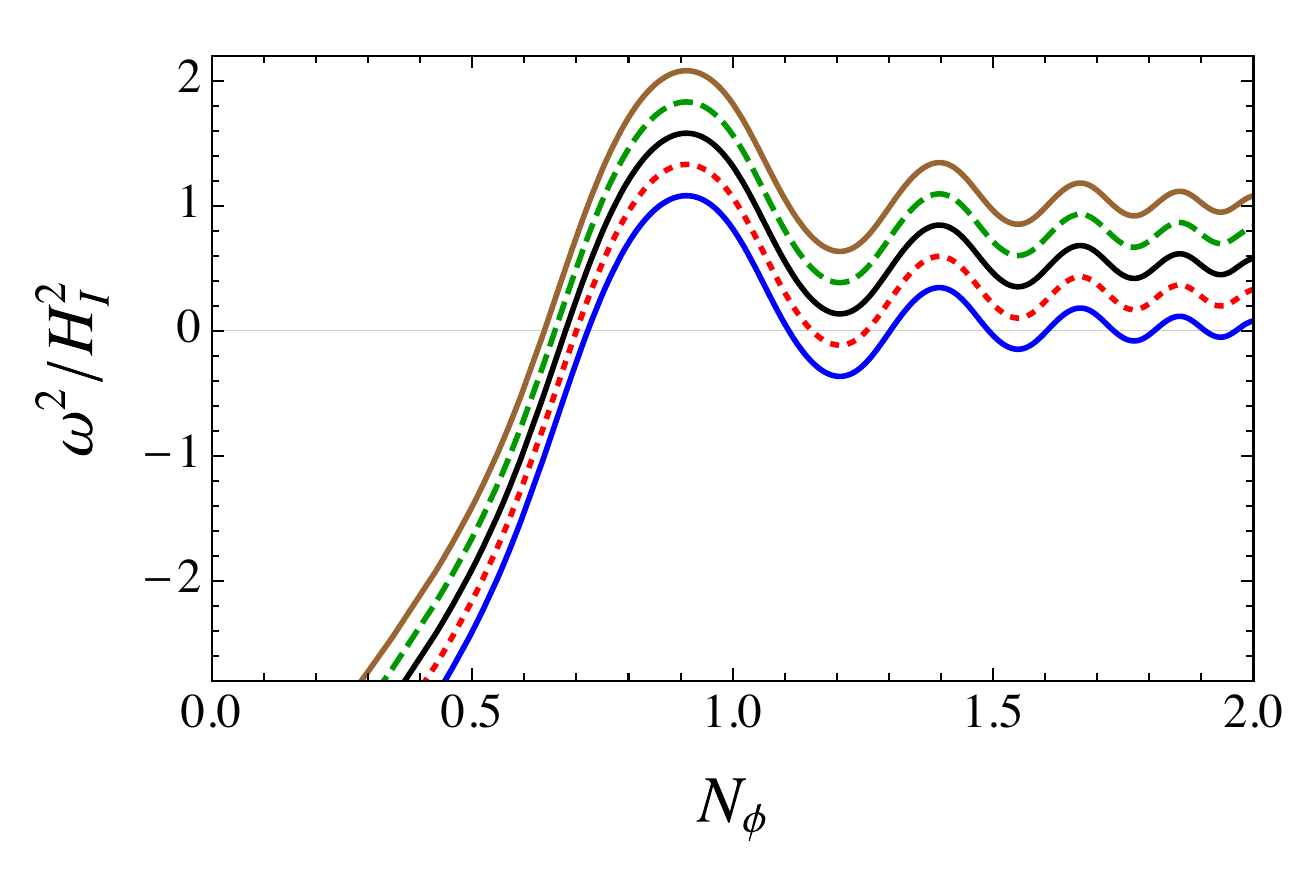}
	\caption{ In the case of gauge preheating, we plot the square of the effective frequency (in units of $H_I^2$) of the dominant gauge field mode as a function of the number of $e$-folds $N_\phi$ before and after the end of inflation (the end of inflation is here taken to be at $N_\phi = 0$). Tachyonic particle production takes place when $\omega^2<0$.  We fix ${\alpha / f} = 9\mP^{-1}$ and we set the effective coupling $M/H_{I} = 0, 0.25 , 0.5, 0.75, 1$ (blue, red-dotted, black, green-dashed and brown lines respectively; increasing values of  $M/ H_{I}$ correspond to curves higher up in the plot). The regions before and after inflation are plotted separately for reasons of visual clarity. Upper panel: first tachyonic burst. Lower panel: parametric resonance after the first zero-crossing of $\omega^2$.}	
	\label{fig:omega2}
\end{center}
\end{figure}


Fig.~\ref{fig:gauge_time} shows the time evolution of the gauge field energy density $\rho_A$ relative to the inflaton energy density $\rho_\phi$ for several values of the coupling $\alpha/ f$ and the gauge field mass. We see that the behavior expected from the evolution of the effective mass squared is verified by the numerical calculation: the initial tachyonic burst shows little deviation for massive and massless gauge bosons, while the subsequent parametric resonance is severely suppressed or completely shut off for sufficiently large mass values.
\begin{figure}[h!]
\begin{center}
	\includegraphics[width=\linewidth]{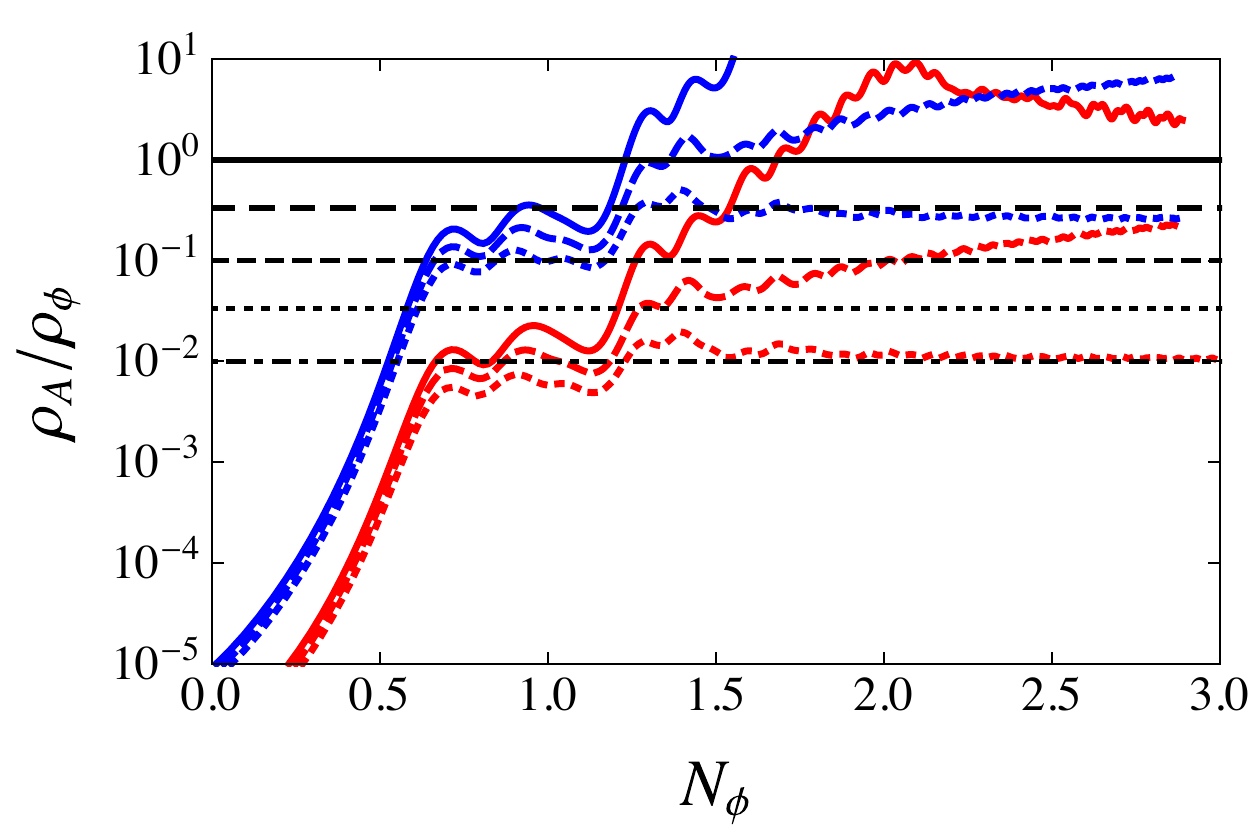}
	\caption{ The ratio of the energy-density in the gauge field to the energy density in the background inflaton field as a function of the number of $e$-folds after the end of inflation $N_\phi$, for $\alpha\mP / f = 9 ,10$ (red and blue respectively) and $M/H_I=0,1,2$ (solid, dashed and dotted lines respectively).
The background equations are solved numerically and the gauge field excitations are computed using Eq.~\eqref{eq:pregauge}, thus neglecting backreaction effects. The transition from preheating to parametric resonance takes place at the location of the first bump.  We define (p)reheating to end once $\rho_\phi = \rho_A$. The horizontal black lines correspond to $\rho_A / \rho_\phi = 1, 0.3, 0.1, 0.03, 0.01$.}
	
	\label{fig:gauge_time}
\end{center}
\end{figure}

As a way to gauge the efficiency of the tachyonic and parametric resonance, we calculate the number of $e$-folds after inflation $\Nr$ for which the energy density of the gauge fields is equal to the energy density of the background inflaton, $\rho_{\rm A} = \rho_{\phi}$. The calculation is performed neglecting backreaction, hence in our approximation $\rho_{\phi}$ denotes the total energy density of the Universe. While this is not a proof of complete preheating, it is a very strong indication of it. We stop the calculation at three $e$-folds after the end of inflation, since all interesting dynamics for the vast majority of cases is contained within that period.

Fig.~\ref{fig:gaugecontour} shows the number of $e$-folds after the end of inflation at which the energy density of the gauge fields first equals the inflaton energy density as a function of the inflaton-gauge coupling $\alpha/f$ and the mass of the gauge field normalized by the Hubble scale at the end of inflation. We can again distinguish the intermediate and large coupling regimes. For intermediate couplings $\alpha/f \lesssim 9.8\,\mP^{-1}$, when the first tachyonic burst is not strong enough and parametric resonance is needed to preheat the Universe, increasing the gauge field mass makes the parametric resonance less effective and smoothly shuts off preheating, before the inflaton can transfer all its energy into the gauge fields. 

In the large coupling regime (for massless gauge bosons) the first tachyonic burst is efficient enough to allow for the majority or all the inflaton energy to be transferred into gauge fields. In this case, introducing a small gauge field mass $M < H_I$ does not significantly affect the early tachyonic burst. It would however suppress any subsequent parametric resonance. Once the gauge field mass is increased beyond the point when it does affect the early tachyonic burst, the subsequent parametric resonance is so far suppressed, that it is completely ineffective. 

Hence, the contour plot of Fig.~\ref{fig:gaugecontour} clearly separates between the intermediate coupling regime $\left ( 9\,\mP^{-1} \lesssim \alpha/f\lesssim  10\, \mP^{-1} \right )$, where an increase in the gauge field mass smoothly leads to a prolonged preheating stage until it is completely shut off, and the large coupling regime $\left (\alpha/f\gtrsim 10 \,\mP^{-1} \right )$, where the gauge field mass either does not appreciably affect the preheating process, or it completely shuts it off.

It is instructive to take a few vertical cuts through the contour plot of Fig.~\ref{fig:gaugecontour} and closely look at how the duration of preheating $\Nr$ changes as a function of the gauge field mass $M$. We plot $\Nr$ as a function of the gauge mass $M$ in the upper panel of Fig.~\ref{fig:preheatefolds} for four different values $\alpha \mP/f = 9, 9.5,10,10.5$. The plot shows a clear difference in the behavior of the $\Nr(M)$ function between the intermediate and the large coupling regimes.

For intermediate couplings $\alpha \mP/f = 9, 9.5$, red and blue lines in the upper panel of Fig.~\ref{fig:preheatefolds}, an adequate fit to the results is a function of the form $\Nr = a+{b/( c-M)}$. Nevertheless, the fit does not capture the steps featured in the functions, which are a consequence of our prescription for defining the end of preheating as the time when $\rho_A=\rho_\phi$. The lower panel of Fig.~\ref{fig:preheatefolds} shows the behavior of the energy density in the gauge bosons $\rho_A$ as a function of $N_\phi$ for the same value of the parameter $\alpha \mP/f$ as in the upper panel, while the black line shows the evolution of the inflaton energy density $\rho_\phi$. Since $\rho_\phi$ and $\rho_A$ are both oscillating functions, the time when the preheating condition is satisfied jumps as a function of $M/H_I$. This behavior is analogous to what we discussed in the middle panel of Fig.~\ref{fig:energydensity}.
 
For larger couplings, the function $\Nr(M)$ has more pronounced features. For $\alpha \mP/f =10$, green line in the upper panel of Fig.~\ref{fig:preheatefolds}, the duration of preheating is almost constant for $M \lesssim 1.4 H_I$, at which point successful preheating abruptly ceases. To explain this behavior, we refer to Fig.~\ref{fig:omega2} and to the lower panel of Fig.~\ref{fig:preheatefolds}. By increasing the gauge field mass enough to suppress the initial tachyonic burst, we have made the parametric resonance completely absent, hence there is no gradual increase of $\Nr$, but rather an abrupt transition from complete to unsuccessful preheating. The case $\alpha \mP/f =10.5$ (brown line) shows even more features. For $M \lesssim 0.7 H_I$ the Universe preheats completely at $\Nr \approx 0$. At $M \sim 0.7 H_I$ there is a discontinuity of $\Delta \Nr \approx 0.3$. The reason is clearly shown by the brown curve of the lower panel of Fig.~\ref{fig:preheatefolds}. The initial tachyonic burst becomes marginally inadequate to preheat the Universe at $M \sim 0.7 H_I$, but the subsequent parametric resonance is still effective. The second feature arises at $M \sim 2.8 H_I$, where we see from the orange curve of the lower panel of Fig.~\ref{fig:preheatefolds} that $\rho_A(N_\phi)$ evolves almost parallel to $\rho_\phi(N_\phi)$ for $N\gtrsim 1.5$. The marginal cases would require a full lattice simulation, in order to determine whether the Universe completely preheats. While the boundaries of Fig.~\ref{fig:gaugecontour} will in general shift, the qualitative features are expected to survive, even if the system is properly simulated.
\begin{figure}[h!]
\begin{center}
	\includegraphics[width=\linewidth]{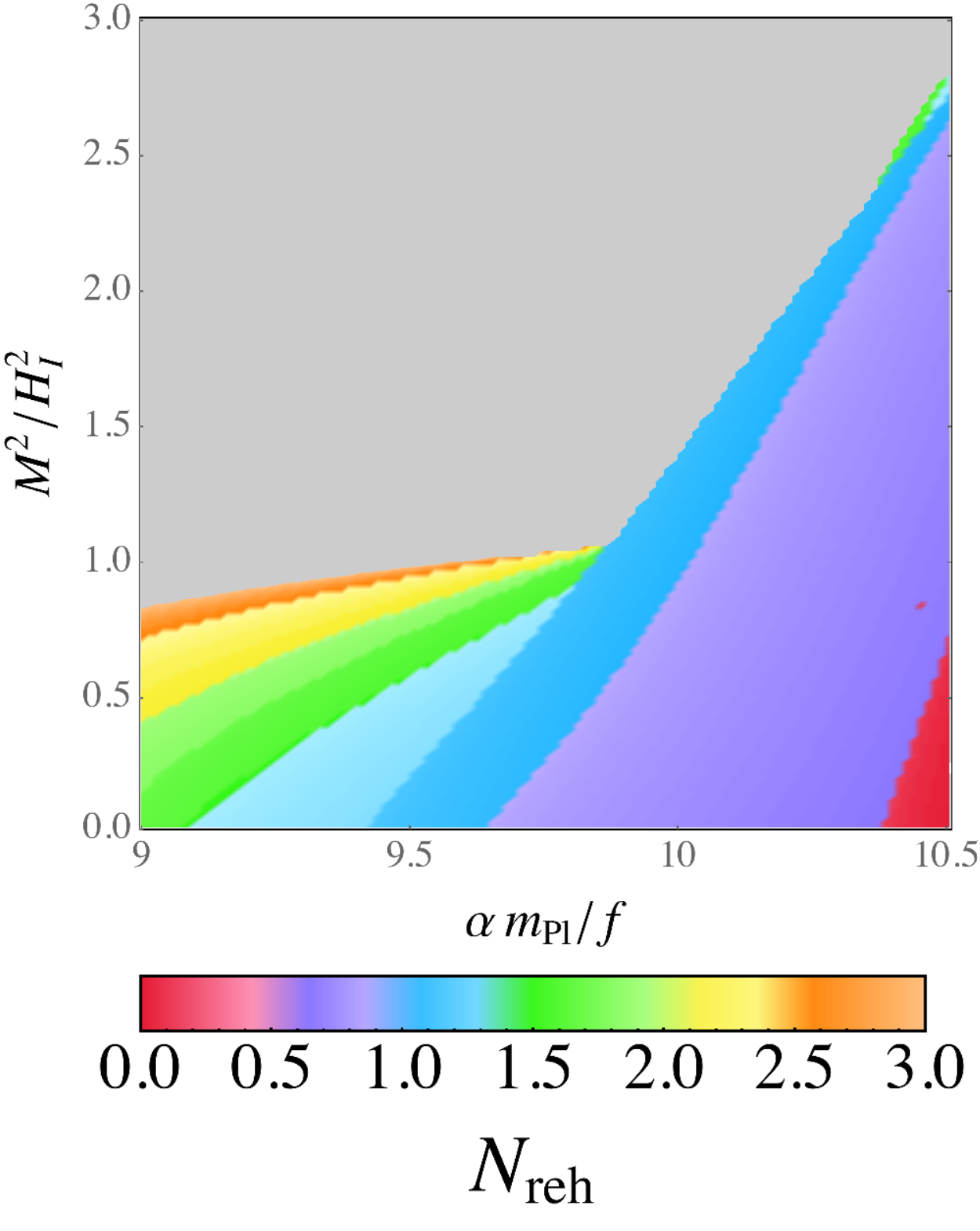}
	\caption{For the case of gauge preheating: the number of $e$-folds after inflation $\Nr$ for which  preheating is complete as a function of the gauge field mass squared (in units of $H_I^2$) and the axion-gauge coupling $\alpha \mP/f$. 
Complete preheating is defined as the time at which $\rho_A = \rho_\phi$, where the gauge field excitations are computed neglecting backreaction.
	In the grey region, the system does not complete preheating.}
	\label{fig:gaugecontour}
\end{center}
\end{figure}
%

\begin{figure}[h!]
\begin{center}
	\includegraphics[width=0.95\linewidth]{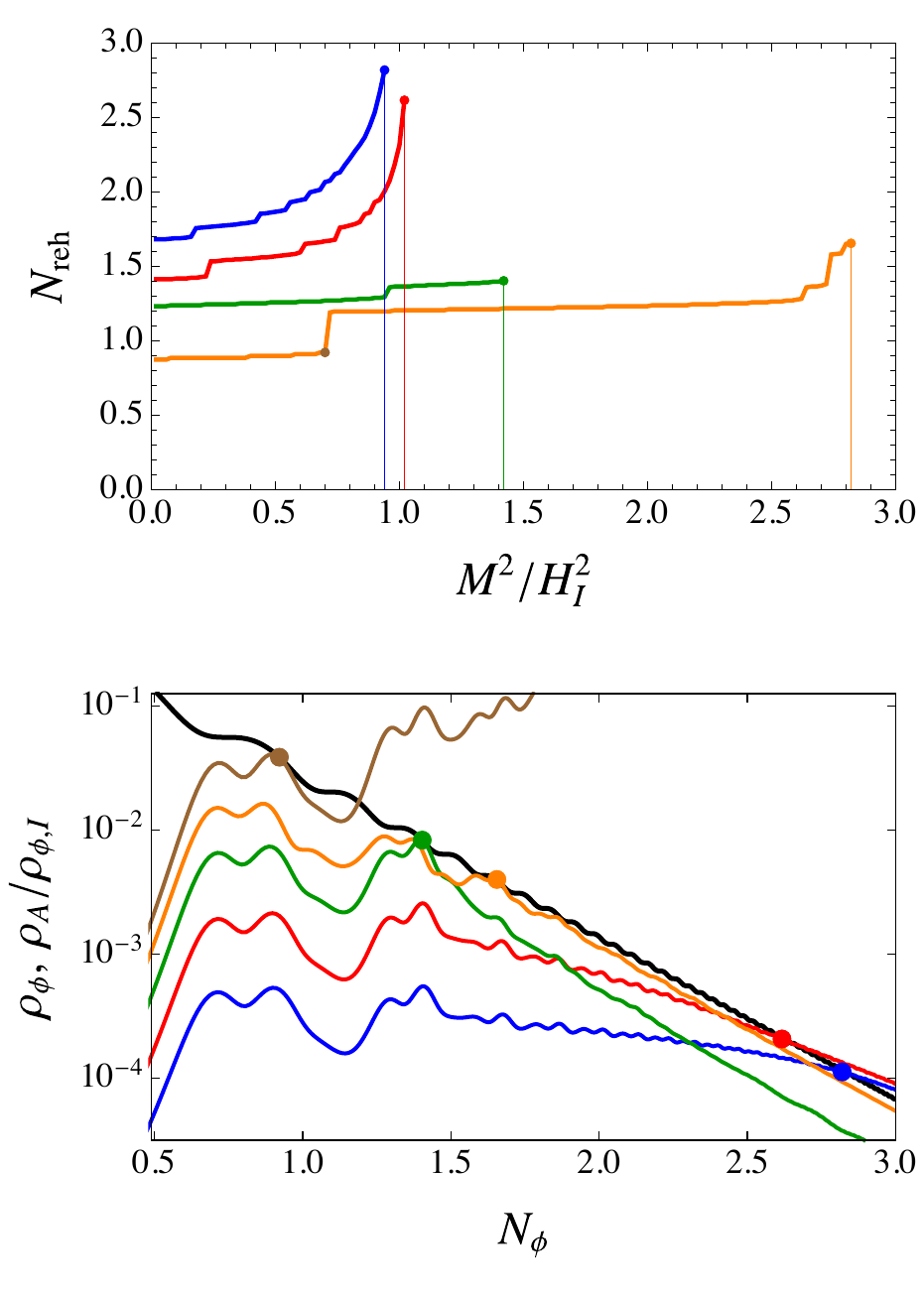}
	\caption{ Upper panel: The time (in $e$-folds) of complete preheating (defined as $\rho_A=\rho_\phi$) for $\alpha \mP/f = 9,9.5,10,10.5$ (blue, red, green and brown lines respectively). The thin vertical lines are introduced for visual clarity to show the values of $M/H_I$ above which preheating fails.
	Lower panel: The evolution of the energy density in gauge fields for parameters denoted by the dots in the upper panel, following the same color-coding. The black curve shows the energy density of the background inflaton field $\rho_\phi$. The dots in the lower panel denote the time of complete preheating in our approximation ($\rho_A=\rho_\phi$).
	}
	\label{fig:preheatefolds}
\end{center}
\end{figure}

Before concluding this section, we want to examine the approximation $M \propto H_I$. While this is a very good approximation after inflation, due to the delayed decay of the Higgs condensate, the exact dynamics of the Higgs VEV during inflation is dependent on its exact potential and must take into account the running of the Higgs self-coupling and any new physics that contributes to this running between the TeV and the inflationary energy scale. We performed a sample of our calculations by allowing the gauge field mass to vary during inflation and the results showed no significant change.  On the qualitative level, the distinction between gradual and sharp shut-off of preheating remains and the quantitative changes on the exact preheating time were smaller than the inherent uncertainties of using a linearized analysis to probe non-linear physics.

\subsection{Abelian vs non-abelian preheating}
\label{sec:abelianvsnonabelian}

Since we are using an abelian $U(1)$ sector as a proxy for preheating into Standard Model Bosons, namely into the non-abelian electroweak gauge bosons, it is worth discussing the possible differences between our study and the full Standard Model.
In general non-Abelian effects are not important for small gauge self-coupling or small values of the gauge-fields themselves. 

In the initial stages of preheating, as long as the linear analysis of the gauge field mode-functions is valid (meaning that the amplitude of the gauge field modes is small), a non-abelian system can be thought of as being de-composed into identical copies of independent abelian fields. An example of that is shown in Ref.~\cite{Adshead:2017xll}, where the preheating behavior of an $SU(2)$ field coupled through a dilaton coupling to the inflaton was simulated.

 {However, once the the gauge field modes become sufficiently populated, then their true non-Abelian nature cannot be neglected. The relevant term in the non-abelian Lagrangian is
\begin{equation}
{\cal L}_{\rm non-abelian} \subset -{1\over 4} f_{abc}f_{ade} A^{b\mu} A_{\mu}^d A^{c\nu}A_{\nu }^e
\end{equation}
where $f_{abc}$ and $f_{ade}$ are $SU(2)$ structure constants.
In the equation of motion for $A_i$, this term in the Lagrangian will induce a term of the form $g^2 A_j A^j A_i$, which can be thought of as an effective non-Abelian mass term. Using a Hartree-type approximation we can define the non-Abelian contribution to the gauge field mass-squared as $m^2_{\rm non-abelian} \sim g^2 \langle A A\rangle $.
This should be added onto the gauge field mass-squared arising from the Higgs condensate.} In general $m_{\rm non-abelian}\gg M$ for the cases when 
there is efficient gauge field production (hence the system nears complete preheating), leading to a further suppression of non-abelian gauge preheating for SM-like systems, compared to our estimates based on Abelian physics. This intuition was verified through lattice simulations in Ref.~\cite{Adshead:2017xll}, where the authors saw a suppression of parametric resonance for increased non-abelian gauge couplings.  

A further phenomenon during preheating of a non-abelian Higgsed sector is described in Ref.~\cite{Enqvist:2015sua}. There, the decay of the Higgs condensate through resonant decay of electroweak bosons is simulated. In this context, the interactions between the non-abelian bosons were found to lead to an extended momentum distribution. In particular, $W$ bosons can scatter and annihilate into $Z$ bosons with significantly higher momenta, than the ones produced through parametric resonance, due to the oscillations of the Higgs condensate. Particles with such high momenta are energetic enough to scatter off the Higgs condensate and fragment it, thereby shutting off any further parametric resonance. 

 {
Fragmentation of a scalar condensate after inflation is a highly non-linear process. It can be triggered either through self-resonance~\cite{Lozanov:2017hjm}
or through scattering of other fields off the condensate. The result is the break-up of a coherently oscillating spatially homogeneous condensate into spatially localized fragments. Once fragmentation occurs, parametric resonance cannot proceed, since the homogeneous oscillating ``pump" is absent. It is possible that the fragmented field evolves into stable or semi-stable localized configurations (such as oscillons~\cite{Amin:2010dc}), which can locally lead to parametric resonance, but the wavelengths involved will be much smaller, typically smaller than the extend of the localized oscillating scalar field fragment.}

In our situation, we do not expect UV modes of the produced gauge fields to fragment the inflaton condensate. However, their effect on the Higgs condensate is unclear. If they act similarly to Ref.~\cite{Enqvist:2015sua}, the Higgs condensate might fragment earlier than expected, reducing the blocking effect of large gauge field masses. 

Finally, we must comment on the possible thermal effects of a gauge boson bath. Since at the energy densities and momenta of the gauge bosons produced during preheating in natural inflation (see for example Ref.~\cite{Adshead:2016iae}), are such that make their interaction rate much faster than the Hubble scale, a gauge boson bath can thermally restore the electroweak symmetry, forcing the Higgs field to roll to its minimum earlier. This would make the gauge fields massless, opening the parametric resonance channel. 
Overall, the extension to a full study of the electroweak sector of the Standard Model would include the simulation of two scalar condensates, the inflaton and the Higgs, along with the gauge bosons, abelian and non-abelian. 

 {
Considering all relevant non-abelian effects, we expect the most dominant to be the extra suppression arising from the increased gauge field mass, due to non-abelian four-point interactions. Hence, we believe our results to represent an upper limit of the effectiveness of tachyonic resonance for Higgsed gauge fields coupled to the inflaton through a Chern-Simons term.}

\subsection{Effects on the reheat temperature}

For models of inflation with coupling through higher dimensional operators as in Eq.~\eqref{eq:V_A}, in many cases (e.g. cosine natural inflation) the requirement for sufficient inflation requires $f\sim \mP$. In that regime, the potential during reheating can roughly be approximated as a simple quadratic $V_\phi \sim m_\phi^2 \phi^2/2$ with $m_\phi \sim 10^{13}\,$ GeV in order to achieve the correct density perturbations.
As discussed in section~\ref{sec:reheatTemp}, more realistic models of inflation which are allowed by Planck data, such as axion monodromy, reduce to a quadratic potential close to to their minimum.
As has been shown in Ref.~\cite{Adshead:2015pva}, the preheating behavior of quadratic and axion monodromy models are equally efficient, for properly chosen values of parameters. 

Then, for a number of $e$-folds $N_\phi$ after the end of inflation,
\begin{align}
	\rho_\phi \approx 0.73\,\rho_{\phi,I}\, e^{-3N_\phi}\, ,
	\label{eq:rhophi}
\end{align}
where $\rho_{\phi, I} = 3\mP^2H_I^2/8\pi$. A simple analytical calculation of the energy density at the end of inflation would give $\rho_\phi (N_\phi=0) = \rho_{\phi,I}$, hence Eq.~\eqref{eq:rhophi} is off by 35\% at $N_\phi = 0$, while being very accurate for $N_\phi\gtrsim1$.
The resulting reheat temperature is then
\begin{align}
	\TRH =\left ( {\rho_\phi \over \sigma_{SB}}\right )^{1/4}  \approx e^{-3{N_{\rm reh}}/4}	10^{16} \,{\rm GeV}
	\label{eq:trh_for_gauge_fields}
\end{align}
where $\sigma_{SB}$ is the Stefan-Boltzmann constant. Reading off the value $\Nr$ for which we obtain a successful reheating from Fig.~\ref{fig:gaugecontour}, we can compute the resulting reheat temperature. 

To illustrate our results, we present numbers for the case $\alpha/f = 9.5 \mP^{-1}$. Without Higgs blocking the number of reheating $e$-folds is $N_{\rm reh} = 1.4$ and, using Eq.~\eqref{eq:trh_for_gauge_fields}, the reheat temperature is $\TRH = 3.5 \times 10^{15}$ GeV.  With Higgs blocking with $M/H_I = 1$ (the maximum value that still allows preheating to complete), we find $N_{\rm reh} = 2.6$ and $\TRH = 1.4 \times 10^{15}\,$GeV. In short, if preheating is successful but delayed, the shift in reheat temperature is less than an order of magnitude, comparable to the changes we previously found for perturbative reheating with Yukawa couplings in section~\ref{sec:perturbative}.
Furthermore, assuming a fixed $m_\phi / H_I$, the change in $N_{\rm reh}$ and relative change in $T_{\rm reh}$ are insensitive to the exact value of $m_{\rm \phi}$.

Far more interesting is the case where preheating is shut down completely due to Higgs blocking. For the case just discussed with $\alpha/f = 9.5 \mP^{-1}$, the complete blocking of preheating happens for $M/H_I \gtrsim 1$. When preheating is completely shut down, then there can be huge differences in the reheat temperature compared to the case of no blocking.  The effect of the Higgs blocking during reheating is largest for these cases.

\subsubsection{Perturbative reheating in the case of derivatively coupled inflaton field} 
As we showed in Fig.~\ref{fig:gaugecontour}, for sufficiently large gauge boson masses the inflaton is unable to transfer all of its energy to the gauge fields and resonant preheating is not sufficient to end inflation. Instead, in those cases, models with a Chern-Simons coupling may still reheat via perturbative decay into gauge fields (see for example Ref.~\cite{Agashe:2014kda}). Specifically we will consider the decay of a massive inflaton $\phi$ into massless gauge bosons $A$ with a decay rate
\begin{align}
	\Gamma_{\phi \to AA} = {\alpha^2 m_\phi^3 \over 64 \pi f^2} \, .
	\label{eq:toAA}
\end{align}
Previously, in section~\ref{sec:perturbative}, we did careful numerical calculations of perturbative reheating. Here, we may do simple analytic estimates instead (and have checked our results are consistent with our numerical calculations assuming a constant decay width, $\Gamma_0$).

In inflation models for which $f\sim \mP$, the perturbative decay rate in Eq.~\eqref{eq:toAA} is so slow (because of the larger factor of $f$ in the denominator) that the decay of the inflaton occurs much later than the decay of the Higgs, and therefore Higgs blocking does not further change the reheat temperature in cases where preheating fails to successfully complete. Thus, one may use standard results in obtaining the reheat temperature (for example, see Ref.~\cite{Kolb:1990vq}). 

For the simple massive inflaton case that we considered, the reheat temperature for perturbative reheating to gauge fields is given from Eq.~\eqref{eq:Trh0} with the decay rate in Eq.~\eqref{eq:toAA},
\begin{align}
	\TRH \approx 9 \times 10^8 \left (  {100\over g_*} \right )^{1/4} \left ( {\alpha \,  \mP \over 10\, f  }\right )\,{\rm GeV }  \, .
	\label{eq:Tr_gauge_pert}
\end{align}

The above results for the reheat temperature can be easily translated into a computation for the time of reheating in $e$-folds after the end of inflation. It is known that assuming a matter-dominated equation of state for the Universe between the end of inflation and the onset of radiation domination, the scale-factor evolves as (see for example Refs.~\cite{Giudice:2000ex, Chung:1998rq})
\begin{align}
	{a_{\rm reh} \over a_I} = \left ({H_I \over H_{\rm reh}} \right )^{2/3} \, .
\end{align}
Using the approximate equalities $H_I \sim m_\phi/2$ and $H_{\rm reh} \sim \Gamma_{\phi \to AA}{/3}$, the above relation becomes $(a_{\rm reh}/  a_I ) \sim (96\pi)^{2/3}(f/\alpha m_\phi)^{4/3}$, leading to $N_{\rm reh} \sim 19$ $e$-folds after the end of inflation. Comparing that to $N_{\rm reh} \lesssim 3$ for tachyonic preheating, we see that the difference can indeed be dramatic.
Finally, note that $T_{\rm reh}$ scales differently with $m_{\phi}$ for perturbative and non-perturbative decays, specifically $T_{\rm reh} \sim {m_\phi}^{1/2}$ for preheating and $T_{\rm reh} \sim {m_\phi}^{3/2}$ for perturbative decays, making the ratio $T_{\rm reh}/T_{\rm reh}^0 \sim m_\phi$ in the case where $M=0$ leads to complete preheating (with $T_{\rm reh}^0$) and a large-enough $M$ requires perturbative decays to reheat (with $T_{\rm reh}$). 

 {
\subsubsection{Gauge fields to fermions}
In order to properly define the reheat temperature in the case of complete preheating, the universe must transition into a thermal plasma consisting of standard model particles. This can be done through decays of gauge bosons into fermions, or through scatterings of gauge bosons into Higgs bosons and fermions. Based on the fact that the gauge fields have masses that equal the Hubble scale, one might think that the decay into fermions will be very fast (see e.g. Refs.~\cite{felder1999, bezrukov2009, garciabellido2009, allahverdi2011}), perhaps enough to suppress or delay tachyonic resonance. A similar discussion regarding the effects of perturbative gauge boson decay on the resonant decay of the Higgs condensate can be found in section~\ref{sec:bosonstofermions}. As in the case of Higgs condensate decay, a simple calculation of the decay of gauge bosons into the lightest fermion species gives
\begin{equation}
	\Gamma_{A \to \bar\psi \psi} = \frac{g'^2\, M}{48\pi} 
	\end{equation}
where $M \sim H$ and $g' \sim 1$. Hence, since $m_\phi \sim H$ as well, the ratio $\Gamma_{A \to \bar\psi \psi} / m_\phi$ is smaller than unity by one or two orders of magnitude and, unlike the case of the Higgs condensate, the effects of perturbative gauge boson decay can be safely ignored in our analysis of tachyonic resonance. 
Scatterings however will be much more efficient, once the universe is dominated by gauge fields\footnote{The effect of gauge boson scattering during the tachyonic regime is an interesting topic in itself, one which we leave for future work.}, as shown in Ref.~\cite{Adshead:2016iae}. We can thus use the energy density of the universe at the end of complete preheating to compute the temperature of the resulting Standard Model plasma.
}

\subsubsection{Distribution of reheat temperatures}
{We have seen how introducing a Higgs-induced mass for the gauge fields can increase the $e$-folds of reheating  and correspondingly reduce the reheat temperature, even by several orders of magnitude in the case when preheating becomes inefficient. In reality however, the effective VEV of the Higgs field will have different values in different patches of the Universe at the end of inflation, resulting in correspondingly different gauge field masses in different patches.
We can study the} resulting distribution of {Hubble} patches with different values of the reheat temperature through the PDF $\tilde f(\TRH)$ as defined in Eqs.~\eqref{eq:TrhCDF} and \eqref{eq:TrhPDF}. The measure of Eq.~\eqref{eq:TrhPDF} can be expressed as
\begin{align}
	\left |  {dh\over dT}  \right | =\left [ \frac{3h}{2H_I} \sqrt{\left(\frac{45}{4\pi^3g_*}\right)^{1/2}\frac{\mP}{H_I}}  e^{-3{N_{\rm reh}}/4}  {d{N_{\rm reh }}\over d\left ({h^2/ H_I^2}\right )} \right]^{-1}
\label{eq:measure_gauge}
\end{align}
where the derivative $d{N_{\rm reh}} / d(h^2/H_I^2)$ can be computed from the data shown in the upper panel of Fig.~\ref{fig:preheatefolds} and we have chosen the value of the gauge coupling $g'=1$ for concreteness.
Fig.~\ref{fig:gaugeTR} shows the reheat temperature as a function of the gauge boson mass, as well as the PDF of the reheat temperature $\tilde f\left (\TRH \right)$. 
We see from Eq.~\eqref{eq:measure_gauge} that the measure $dh/dT$ diverges at $h\to 0$, hence each PDF in the lower panel of Fig.~\ref{fig:gaugeTR} peaks at the maximum possible reheat temperature for that particular choice of the axion-gauge coupling.
We must note that, as shown in Fig.~\ref{fig:gaugeTR}, the PDF's are not normalized to unity. The probability that is not enclosed in each PDF of Fig.~\ref{fig:gaugeTR} is concentrated in a delta function contribution at the reheat temperature given by Eq.~\eqref{eq:Tr_gauge_pert}. This can be also understood from the upper panel of Fig.~\ref{fig:gaugeTR}: for values of the gauge field mass higher than the ones plotted here for each of the couplings, reheating proceeds perturbatively and the temperature is lower by several orders of magnitude $\TRH \sim 10^{{9}}\, {\rm GeV}$. 
Finally, the exact shape of the PDF's is  sensitive to the measure given in Eq.~\ref{eq:measure_gauge}. Due to the discrete set of values used for $M/H_I$, the function $N_{\rm reh}(h/H_I)$ exhibits discontinuities, as shown in Fig.~\ref{fig:preheatefolds}. In order to compute the derivative $dN_{\rm reh}/ d(h^2/H_I^2)$, we (piece-wisely) fitted the curves in the upper panel of Fig.~\ref{fig:preheatefolds} by polynomials, which we then differentiated. The choice of the polynomial form used affects  the local features of the PDF's shown in Fig.~\ref{fig:gaugeTR}, but not their range and general form.
\begin{figure}[h!]
\begin{center}
	\includegraphics[width=\linewidth]{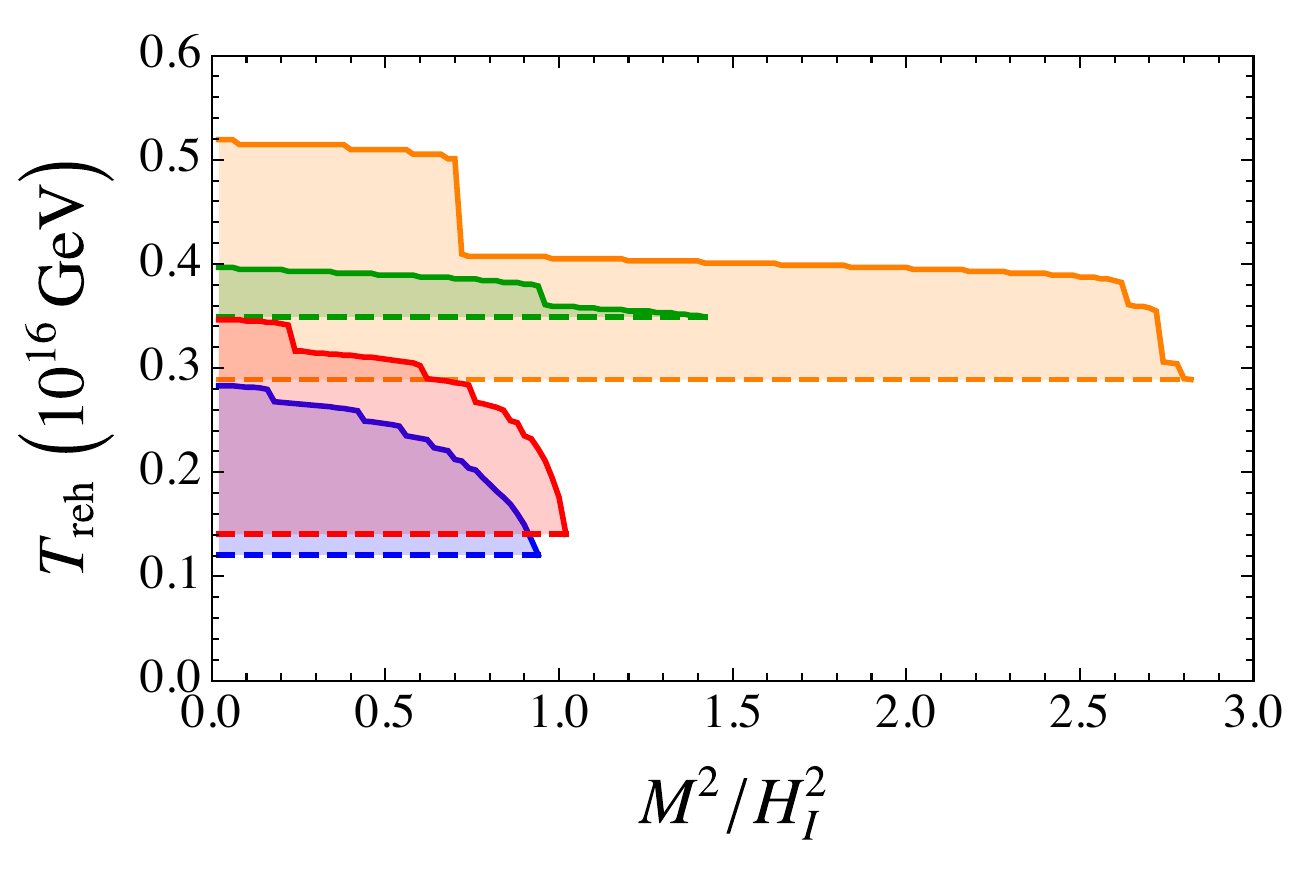}
	\\
	\medskip
	\includegraphics[width=\linewidth]{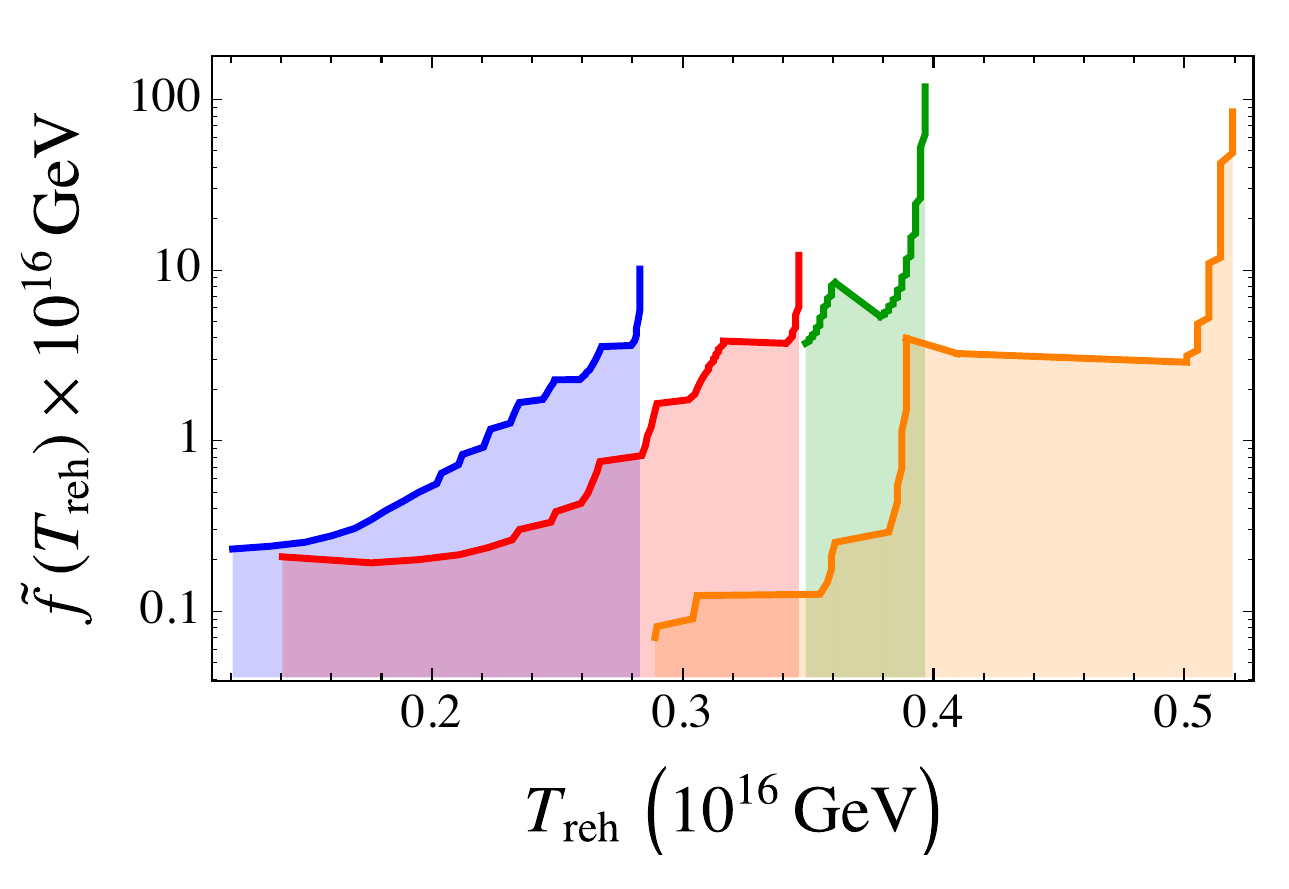}
	\caption{Upper panel: Reheat temperature as a function of the gauge field mass for $\alpha \mP/ f = 9,9.5,10,10.5$ (blue, red, green and orange respectively) {considering a quadratic inflaton potential}.
	Lower panel: The Probability Density Function as a function of reheat temperature for the same color-coding as the above panel. Each PDF also includes a delta-function component at $\TRH \sim 10^{{9}}\, {\rm GeV}$ arising from the perturbative decays. {If one removes the Higgs-induced gauge boson mass terms, the PDF collapses to a Dirac function centered at the right extremal value of each of the colored curves.} Note that, assuming a fixed $m_\phi / H_I$, the relative change in $T_{\rm reh}$ due to the effects of Higgs blocking are insensitive to the exact value of $m_{\rm \phi}$.		}
	\label{fig:gaugeTR}
\end{center}
\end{figure}

Summing up, we have seen that the introduction of Higgs-induced gauge field masses can change the (p)reheating behavior of models with Chern-Simons interactions in ways that range from mild to dramatic. 
For intermediate values of axion-gauge couplings, as one increases the gauge masses, preheating is prolonged until it smoothly shuts off. However, for large values of axion-gauge couplings, gauge field masses either leave the reheating behavior almost unaffected or completely suppress it. The resulting distribution of reheat temperatures in different Hubble patches is split into two pieces: an extended one, arising from non-perturbative preheating, and a delta-function-shaped one, due to perturbative inflaton decays.

\section{Discussion and Conclusions} \label{sec:discussion}

In this work we examined the effects of a non-zero Higgs VEV during inflation on the reheating history of the Universe. 
  We studied the effects on the multiple stages of reheating: perturbative decays from coherent oscillations of the inflaton field as well as resonant particle production (preheating).
  We studied the quite general case of the inflaton decays to Standard Model fermions with coupling through Yukawa interactions.   Furthermore, we studied Higgs effects in the case of Abelian Gauge fields coupled to the inflaton through a Chern-Simons coupling term. The latter arises generically in models of natural inflation, where the inflaton respects a (softly broken) shift symmetry, which severely restricts its possible couplings to ones containing derivative interactions. We used this as a toy model for the electroweak sector of the SM and studied preheating in this case. The perturbative decays can be thought of as a more generic case, while in the  case of resonant particle production due to the Chern-Simons couplings, Higgs effects can have a more dramatic effect.

In the case of perturbative decay into fermions, the Hubble-sized Higgs VEV induces large SM fermion masses. This can delay reheating for up to four $e$-folds in some cases, compared to the results one derives when neglecting the large fermion masses. This can reduce the reheat temperature by an order of magnitude. It is important to note, that these mass-blocking effects are only relevant for inflation that occurs at a sufficiently high scale.  At low enough energy scale, the Higgs VEV could be low enough (since it is determined by the Hubble scale at the end of inflation) that all the particles that acquire a mass through the Higgs mechanism have masses substantially below the inflaton mass $m_\phi$. Hence for $H_I \lesssim 0.01m_\phi$, which can be the case for a potential that is quadratic near the bottom but becomes concave away from it, the blocking effects described here are largely absent and the standard reheating estimates hold. 

In the case of a Chern-Simons coupling to gauge fields, the gauge-field equations of motion exhibit a tachyonic instability, which allows for a very efficient transfer of energy from the inflaton to the radiation modes. This can allow for an almost instantaneous preheating, leading to reheat temperatures as high as $10^{16}\, {\rm GeV}$ for high-scale inflation. If the gauge fields acquire large masses due to the Higgs mechanism, the tachyonic and parametric resonances can be suppressed. This can even make preheating inefficient, leading to a prolonged reheat stage through perturbative decays resulting in a reheat temperature of $10^{{9}}\,$GeV. An extension to non-Abelian gauge fields can reveal interesting phenomenology, as both the Higgs-induced mass and the non-Abelian self-coupling can suppress preheating  (see for example Ref.~\cite{Adshead:2017xll}). We leave this for future work.

The Higgs blocking effects can have a number of observational consequences. First, any connection of inflationary models to CMB measurements must take into account the duration of the (p)reheating stage, since it affects the expansion history of the Universe, which in turn determines at which point during inflation the CMB-relevant modes exited the horizon. Simply put, Higgs blocking effects can shift the predictions of (especially high-scale) models on the $n_s-r$ plane. The decrease in the number of $e$-folds for which observable modes of inflaton fluctuations can exit the horizon during inflation due to the Higgs blocking effect is given by~\cite{Liddle:2003as} 
\begin{align}
	\Delta N_{\rm hor} \approx  - {1 \over 4} \Delta N_\phi ,
	\label{eq:efolds}
\end{align}
where $\Delta N_\phi$ is the number of extra $e$-folds of matter domination due to delayed reheating. Specifically, for the perturbative inflaton decay via Yukawa couplings that we considered in section~\ref{sec:perturbative}, $\Delta N_{\rm hor}  \lesssim 1$.
For the case of natural inflation with Chern-Simons coupling of the inflaton to gauge fields
discussed in section~\ref{sec:gauge},
the change can be larger.  Higgs blocking can cause preheating to fail, in which case, instead of instantaneous preheating, the resulting change can be as large as $\Delta N_{\rm hor} \simeq 5$.
Thus, as the energy density of the inflaton condensate at the time of reheating decreases due to the effects of Higgs blocking, the shift in the $n_s-r$ plane is towards lower $n_s$ and higher $r$ with a magnitude $\Delta N_{\rm hor} $. 

More interesting than the change in the number of observable $e$-folds is the lowered reheat temperature, which can change by many orders of magnitude due to Higgs blocking for some models. One consequence of the lower reheat temperature is the possible effect on baryogenesis models. Some inflation models have been proposed as origins of the matter/antimatter asymmetry of the Universe via leptogenesis in the inflaton decay that later converts to nonzero baryon number~\cite{Giudice:1999fb, Adshead:2015jza, Kusenko:2014lra, Pearce:2015nga, Yang:2015ida, Alexander:2004us, Caldwell:2017chz}.  In these models, if the reheat temperature is too high, the lepton asymmetry generated during reheating could be wiped out by rapid lepton number violating interactions~\cite{Yang:2015ida, Adshead:2017znw}.  A lower reheat temperature would avoid such processes. 

Furthermore, the stochastic nature of the Higgs VEV during inflation leads to different Hubble patches at the end of inflation acquiring different reheat temperatures. In the case of perturbative decays into fermions, they can vary by an order of magnitude, whereas in the case of the axion-gauge coupling, the temperatures can be as disparate as $10^{16}\,{\rm GeV}$ and $10^{{9}}\,{\rm GeV}$. Depending on the physical processes taking place after inflation, this can lead to both adiabatic and baryon isocurvature fluctuations. 

Adiabatic fluctuations arise because the Universe exhibits a space-dependent reheat temperature, due to the correspondingly space-dependent Higgs-induced particle masses. 
Ref.~\cite{Dvali:2003em} proposed that observable primordial adiabatic fluctuations can be generated during reheating if one postulates a space-dependent decay rate of the inflaton to radiation.  Ref.~\cite{Dvali:2003em} used a specific construction within the context of the Minimal Supersymmetric Standard Model (MSSM) to generate such space-dependent inflaton decays, in which the inflaton decay rate $\Gamma_\phi$ inherited the fluctuations of a light scalar field during inflation. However, Higgs-induced SM masses can have a similar impact on inflaton decays 
without requiring special assumptions regarding the inflaton couplings to SM particles.
A computation of the spectral index and higher-point correlators of the produced adiabatic modes due to Higgs-modulated reheating is an interesting problem in itself, one which we leave for future work.
In any case, 
if one simply wishes to isolate the CMB from such contributions during reheating, the Higgs motion must be restricted during the time, at which the CMB-relevant modes exit the horizon, or the reheating process must proceed in ways that circumvent the Higgs-induced delay.
Adiabatic fluctuations at smaller scales are somewhat less constrained and they can be masked by other effects. For example, massive neutrinos are known to suppress power at small scales~\cite{Jimenez:2016ckl, Palanque-Delabrouille:2015pga, Cuesta:2015iho, Hu:1997mj, Lesgourgues:2006nd}, hence disentangling the extra contributions from Higgs-delayed reheating from possible suppression factors can be challenging.

Furthermore, if the above mentioned inhomogeneity of the reheat temperature is applied to the case of inflationary leptogenesis, it can lead to a space-dependent baryon number. Simply put, different Hubble patches will acquire different reheat temperatures and lepton-violating processes will occur at different rates, leading to space-dependent lepton-number washout, even if lepton-number was produced homogeneously across the observable Universe at or near the end of inflation. Sphaleron processes will in turn convert part of the space-dependent lepton number into a space-dependent baryon number. This can severely alter predictions of models, in which the baryon asymmetry of the Universe can be traced back to a lepton asymmetry generated during inflation. A spatially varying baryon number can lead to baryon isocurvature fluctuations through similar processes, as the ones considered in Ref.~\cite{Kusenko:2017kdr}. There it was argued that small-scale fluctuations in the baryon abundance can lead to differences in the  star formation rates, which can in turn source fluctuations in the Cosmic Infrared Background (CIB). These baryon isocurvature fluctuations will be highly correlated with the above mentioned adiabatic ones, however they will depend heavily on the details of the various inflationary leptogenesis models. While we do not attempt to provide any concrete computation of such effects here, we point out that Higgs-delayed reheating may provide a new way of constraining inflationary leptogenesis models, or equivalently baryon isocurvature may provide novel ways of observing the dynamics of the Higgs field during inflation.

Overall, the fluctuations generated by Higgs-delayed reheating, or the lack thereof, may be used to place constraints on the behavior of the Higgs field during and immediately after inflation. This is a novel probe of electroweak physics, which --even taking into account its inherent complexity-- can be a useful tool in the modern era of precision cosmology. 

\begin{acknowledgments}
We would like to thank Peter Adshead, Gordon Kane, and Yue Zhao for the useful discussions and comments that led to the present work, and the program ``Advances in Theoretical Cosmology in Light of Data" at Nordita for hosting. KF, PS and LV acknowledge support by the Vetenskapsr\r{a}det (Swedish Research Council) through contract No. 638-2013-8993 and the Oskar Klein Centre for Cosmoparticle Physics. KF, PS, and LV acknowledge support from the Department of Energy through DoE grant DE-SC0007859 and the Leinweber Center for Theoretical Physics at the University of Michigan.  KF and LV thank Perimeter Institute, where part of this work was conducted, for hospitality.
The work of EIS was supported in part by NASA Astrophysics Theory Grant NNX17AG48G. EIS gratefully acknowledges support from a Fortner Fellowship at the University of Illinois at Urbana-Champaign, and also the Dutch Organisation for Scientific Research (NWO).
\end{acknowledgments}

\appendix

\section{Abelian Higgsed Model}
\label{sec:abelianhiggsedmodel}

For completeness we include here the derivation of the equations of motion for the gauge fields after the Spontaneous Symmetry Breaking has occurred during inflation. As a toy-model, we only consider a $U(1)$ field coupled to both the inflaton $\phi$ and the complex scalar Higgs field $h$. The full Lagrangian is
\begin{eqnarray}
&&S = \int \sqrt{-g} d^4 x \left \{ 
{\mP^2\over 2}R 
+{1\over 2} \partial_\mu\phi \partial^\mu\phi  -V(\phi) \right .
\\
\nonumber
&&\left .
-{1\over 4}F_{\mu\nu} F^{\mu\nu}- {\alpha\over 4f}\phi F_{\mu\nu}\tilde F^{\mu\nu} 
+ D_\mu h^* D^\mu h - V\left (|h|^2 \right )
\right \} \, ,
\end{eqnarray}
where $D_\mu h = \partial_\mu h + i q A_\mu h$ and $V\left (|h|^2 \right )$ is the Higgs potential during inflation. The higgs field can be written as
\begin{equation}
h = {v+\sigma(x) \over \sqrt{2}} e^{i\theta(x)}
\end{equation}
where $v$ is the Higgs VEV during inflation and $\sigma, \theta$ are real functions. The Higgs kinetic term becomes
\begin{equation}
|D_\mu h|^2  = {1\over 2} (\partial_\mu \sigma)^2 + {(v+\sigma)^2\over 2} \left ( \partial_\mu \theta+q A_\mu\right )^2
\end{equation}
We now re-define the field strength as $A_\mu \to A_\mu - \partial_\mu \theta /q $, so that the above term becomes
\begin{equation}
|D_\mu h|^2  = {1\over 2} (\partial_\mu \sigma)^2 +{v^2 \over 2}A_\mu A^\mu +{\sigma^2\over 2} A_\mu A^\mu + v \sigma A_\mu A^\mu
\end{equation}
where we see the effective mass term ${v^2 \over 2}A_\mu A^\mu $ for the gauge field.
The inflaton field is not directly coupled to the Higgs field, hence the equation of motion is unchanged compared to the results in Refs.~\cite{Adshead:2015pva, Adshead:2016iae}. The equation of motion of the massive part of the Higgs field $\sigma(x)$ is
\begin{equation}
(\partial_\tau^2 + 2 {\cal H} \partial_\tau - \partial_i \partial_i )\sigma + a^2 {\partial V(\sigma) \over \partial\sigma} = -(v+\sigma) A_\mu A^\mu   
\end{equation}
We will not further consider the motion of the $\sigma$ field here. The equation of motion for the gauge field strength $A_\mu$ is
\begin{equation}
\partial_\alpha \left (\sqrt{-g} F^{\alpha\beta} \right ) + {\alpha \over f} \partial_\alpha \left (\sqrt{-g}  \phi \tilde F^{\alpha\beta} \right ) = - {v^2 \over 2}A^\beta \equiv -m^2 A^\beta
\end{equation}
The temporal ($\beta=0$) equation is a generalized version of the Gauss' law constraint
\begin{equation}
\partial_j\partial_j A_0 - \partial_\tau \partial_i A_i + {\alpha \over f} \epsilon_{ijk} (\partial_k \phi )(\partial_i A_j)  -m^2A_0 =0
\end{equation}
and the spatial ($\beta=i$) equations are the actual dynamical equations of motion for the spatial components of the gauge field $A_i$
\begin{eqnarray}
\nonumber
&&-\partial_\tau(\partial_\tau A_i - \partial_i A_0) + \partial_m(\partial_m A_i - \partial_i A_m) 
\\
\nonumber
&&+ {\alpha\over f} \epsilon_{imk} (\partial_\tau \phi) (\partial_m A_k) 
- {\alpha\over f} \epsilon_{imk}(\partial_m\phi)(\partial_\tau A_k - \partial_k A_0)  
\\
&&-m^2 A_i=0
\label{eq:A_ieom}
\end{eqnarray}
Since we want to study the generation of gauge fields from the background inflaton, we set $\phi(\vec x,t) \equiv \phi( t)$, leading to $\partial_m\phi=0$ in Eq.\eqref{eq:A_ieom}. Decomposing the vector field in one longitudinal mode $A^L$ and two transverse mores of left- and right-handed helicity $A^\pm$ the equations of motion become
\begin{eqnarray}
\left ( \partial_\tau^2 + k^2 \mp {\alpha\over f}  k (\partial_\tau \phi)  + m^2 \right)   A^\pm_k &=&0
\\
\left ( \partial_\tau^2 + k^2 + m^2\right )A^L_k &=&0
\end{eqnarray}
We see that the Chern-Simons coupling does not appear in the longitudinal mode. It thus remains conformally coupled and its classical modes are not excited, hence we only consider the transverse modes. This derivation for the transverse gauge field modes thus leads to an effective Lagrangian with the potential given in Eq.~\eqref{eq:V_A}.


\begin{thebibliography}{99}

\bibitem{Guth:1980zm} A.~H.~Guth, Phys.\ Rev.\ D {\bf 23}, 347 (1981).

\bibitem{Linde:1981mu} A.~D.~Linde, Phys.\ Lett.\ B {\bf 108}, 389 (1982).

\bibitem{Albrecht:1982wi} A.~Albrecht and P.~J.~Steinhardt, Phys.\ Rev.\ Lett.\ {\bf 48}, 1220 (1982).

\bibitem{Enqvist:2013kaa} 
  K.~Enqvist, T.~Meriniemi, S.~Nurmi,
  JCAP {\bf 1310}, 057 (2013)
  doi:10.1088/1475-7516/2013/10/057
  [arXiv:1306.4511 [hep-ph]].

  \bibitem{Kusenko:2014lra} 
  A.~Kusenko, L.~Pearce and L.~Yang,
  Phys.\ Rev.\ Lett.\  {\bf 114}, no. 6, 061302 (2015)
  [arXiv:1410.0722 [hep-ph]].

\bibitem{Starobinsky:1994bd} 
  A.~A.~Starobinsky and J.~Yokoyama,
  Phys.\ Rev.\ D {\bf 50}, 6357 (1994)
  doi:10.1103/PhysRevD.50.6357
  [astro-ph/9407016].


\bibitem{Cook:2015vqa} 
  J.~L.~Cook, E.~Dimastrogiovanni, D.~A.~Easson and L.~M.~Krauss,
  JCAP {\bf 1504}, 047 (2015)
  [arXiv:1502.04673 [astro-ph.CO]].


\bibitem{Liddle:2003as} 
  A.~R.~Liddle and S.~M.~Leach,
  Phys.\ Rev.\ D {\bf 68}, 103503 (2003)
  [astro-ph/0305263].

\bibitem{Martin:2006rs} 
  J.~Martin and C.~Ringeval,
  JCAP {\bf 0608}, 009 (2006)
  [astro-ph/0605367].

\bibitem{Lorenz:2007ze} 
  L.~Lorenz, J.~Martin and C.~Ringeval,
  JCAP {\bf 0804}, 001 (2008)
  [arXiv:0709.3758 [hep-th]].

\bibitem{Adshead:2010mc} 
  P.~Adshead, R.~Easther, J.~Pritchard and A.~Loeb,
  JCAP {\bf 1102}, 021 (2011)
  [arXiv:1007.3748 [astro-ph.CO]].
  
\bibitem{Easther:2011yq} 
  R.~Easther and H.~V.~Peiris,
  Phys.\ Rev.\ D {\bf 85}, 103533 (2012)
  [arXiv:1112.0326 [astro-ph.CO]].
  
\bibitem{Dai:2014jja} 
  L.~Dai, M.~Kamionkowski and J.~Wang,
  Phys.\ Rev.\ Lett.\  {\bf 113}, 041302 (2014)
  [arXiv:1404.6704 [astro-ph.CO]].
  
\bibitem{Martin:2014nya} 
  J.~Martin, C.~Ringeval and V.~Vennin,
  Phys.\ Rev.\ Lett.\  {\bf 114}, no. 8, 081303 (2015)
  [arXiv:1410.7958 [astro-ph.CO]].
  
\bibitem{Drewes:2015coa} 
  M.~Drewes,
  JCAP {\bf 1603}, no. 03, 013 (2016)
  [arXiv:1511.03280 [astro-ph.CO]].
  
  \bibitem{Feng:2003nt} 
  B.~Feng, X.~Gong and X.~Wang,
  Mod.\ Phys.\ Lett.\ A {\bf 19}, 2377 (2004)
  [astro-ph/0301111].
    
  \bibitem{Giudice:1999fb} 
  G.~F.~Giudice, M.~Peloso, A.~Riotto and I.~Tkachev,
  JHEP {\bf 9908}, 014 (1999)
  [hep-ph/9905242].
  
  
  
  

\bibitem{Adshead:2015kza} 
  P.~Adshead and E.~I.~Sfakianakis,
  JCAP {\bf 1511}, no. 11, 021 (2015)
  [arXiv:1508.00891 [hep-ph]].

  
    \bibitem{Adshead:2015jza} 
  P.~Adshead and E.~I.~Sfakianakis,
  Phys.\ Rev.\ Lett.\  {\bf 116}, no. 9, 091301 (2016)
  [arXiv:1508.00881 [hep-ph]].
    
  \bibitem{Pearce:2015nga} 
  L.~Pearce, L.~Yang, A.~Kusenko and M.~Peloso,
  Phys.\ Rev.\ D {\bf 92}, no. 2, 023509 (2015)
  [arXiv:1505.02461 [hep-ph]].
  
  \bibitem{Yang:2015ida} 
  L.~Yang, L.~Pearce and A.~Kusenko,
  Phys.\ Rev.\ D {\bf 92}, no. 4, 043506 (2015)
  [arXiv:1505.07912 [hep-ph]].
  
  \bibitem{Alexander:2004us} 
  S.~H.~S.~Alexander, M.~E.~Peskin and M.~M.~Sheikh-Jabbari,
  Phys.\ Rev.\ Lett.\  {\bf 96}, 081301 (2006)
  [hep-th/0403069].
  
  \bibitem{Caldwell:2017chz} 
  R.~R.~Caldwell and C.~Devulder,
  arXiv:1706.03765 [astro-ph.CO].

\bibitem{Adshead:2017znw} 
P.~Adshead, A.~J.~Long and E.~I.~Sfakianakis,
  arXiv:1711.04800 [hep-ph].

\bibitem{Freese:1990rb} 
  K.~Freese, J.~A.~Frieman and A.~V.~Olinto,
  Phys.\ Rev.\ Lett.\  {\bf 65}, 3233 (1990).

\bibitem{Adams:1990pn}
     F.~C.~Adams, K.~Freese, A.~H.~Guth,
     Phys.\ Rev.\ D {\bf 43}, 955 (1991),
      doi:10.1103/PhysRevD.43.965.

\bibitem{Adshead:2015pva} 
  P.~Adshead, J.~T.~Giblin, T.~R.~Scully and E.~I.~Sfakianakis,
  JCAP {\bf 1512}, no. 12, 034 (2015)
  [arXiv:1502.06506 [astro-ph.CO]].

\bibitem{Adshead:2016iae} 
  P.~Adshead, J.~T.~Giblin, T.~R.~Scully and E.~I.~Sfakianakis,
  JCAP {\bf 1610}, 039 (2016)
  [arXiv:1606.08474 [astro-ph.CO]].



\bibitem{Neronov:1900zz} 
  A.~Neronov and I.~Vovk,
  Science {\bf 328}, 73 (2010)
  [arXiv:1006.3504 [astro-ph.HE]].
  
  \bibitem{Tavecchio:2010mk} 
  F.~Tavecchio, G.~Ghisellini, L.~Foschini, G.~Bonnoli, G.~Ghirlanda and P.~Coppi,
  Mon.\ Not.\ Roy.\ Astron.\ Soc.\  {\bf 406}, L70 (2010)
  [arXiv:1004.1329 [astro-ph.CO]].
  
  \bibitem{Dolag:2010ni} 
  K.~Dolag, M.~Kachelriess, S.~Ostapchenko and R.~Tomas,
  Astrophys.\ J.\  {\bf 727}, L4 (2011)
  [arXiv:1009.1782 [astro-ph.HE]].
  
  \bibitem{Essey:2010nd} 
  W.~Essey, S.~Ando and A.~Kusenko,
  Astropart.\ Phys.\  {\bf 35}, 135 (2011)
  [arXiv:1012.5313 [astro-ph.HE]].
  
  \bibitem{Taylor:2011bn} 
  A.~M.~Taylor, I.~Vovk and A.~Neronov,
  Astron.\ Astrophys.\  {\bf 529}, A144 (2011)
  [arXiv:1101.0932 [astro-ph.HE]].
  
  \bibitem{Takahashi:2013lba} 
  K.~Takahashi, M.~Mori, K.~Ichiki, S.~Inoue and H.~Takami,
  Astrophys.\ J.\  {\bf 771}, L42 (2013)
  [arXiv:1303.3069 [astro-ph.CO]].
  
  \bibitem{Finke:2013tyq} 
  J.~Finke {\it et al.} [Fermi-LAT Collaboration],
  eConf C {\bf 121028}, 365 (2012)
  [arXiv:1303.5093 [astro-ph.HE]].
  
  \bibitem{Kachelriess:2012mc} 
  M.~Kachelriess, S.~Ostapchenko and R.~Tomas,
  J.\ Phys.\ Conf.\ Ser.\  {\bf 375}, 052030 (2012).


\bibitem{Adams:1992bn}
	F.~C.~Adams, J.~R.~Bond, K.~Freese, J.~A.~Frieman, A.~V.~Olinto,
	Phys.\ Rev.\ D {\bf 47}, 426 (1993),
	doi:10.1103/PhysRevD.47.426,
	[arXiv/9207245 [hep-ph]].

\bibitem{Gaillard:1995az}
	M.~K.~Gaillard, H.~Murayama, K.~A.~Olive,
	Phys.\ Lett.\ B {\bf 355}, 71 (1995),
	doi:10.1016/0370-2693(95)00773-E,
	[arXiv/9504307 [hep-ph]].

\bibitem{Kawasaki:2000ws}
	M.~Kawasaki, M.~Yamaguchi, T.~Yanagida,
	Phys.\ Rev.\ D {\bf 63}, 103514 (2001),
	doi:10.1103/PhysRevD.63.103514,
	[arXiv/0011104 [hep-ph]].
	
\bibitem{Banks:2003sx}
	T.~Banks, M.~Dine, P.~J.~Fox, E.~Gorbatov,
	JCAP {\bf 0306}, 001 (2003),
	doi:10.1088/1475-7516/2003/06/001,
	[arXiv/0303252 [hep-th]].

\bibitem{Hsu:2003cy}
	J.~P.~Hsu, R.~Kallosh, S.~Prokushkin,
	JCAP {\bf 0312}, 009 (2003),
	doi:10.1088/1475-7516/2003/12/009,
	[arXiv/0311077 [hep-th]].
	
\bibitem{Hsu:2004hi}
	J.~P.~Hsu, R.~Kallosh,
	JHEP {\bf 0404}, 042 (2004),
	doi:10.1088/1126-6708/2004/04/042,
	[arXiv/0402047 [hep-th]].

\bibitem{Freese:2004un}
      K.~Freese and W.~H.~Kinney,
      Phys.\ Rev.\ D {\bf 70}, 083512 (2004),
      doi:10.1103/PhysRevD.70.083512.

\bibitem{Kim:2004rp}
      J.~E.~Kim, H.~P.~Nilles, M.~Peloso,
      JCAP {\bf 0501}, 005 (2005),
      doi:10.1088/1475-7516/2005/01/005,
      [arXiv:0409138 [hep-ph]]

\bibitem{Degrassi:2012ry} 
 G.~Degrassi, S.~Di Vita, J.~Elias-Miro, J.~R.~Espinosa, G.~F.~Giudice, G.~Isidori, and A.~Strumia,
  JHEP {\bf 1208}, 098 (2012)
  doi:10.1007/JHEP08(2012)098
  [arXiv:1205.6497 [hep-ph]].

\bibitem{buttazzo2013} D.~Buttazzo, G.~Degrassi, P.~P.~Giardino, G.~F.~Giudice, F.~Sala, A.~Salvio, and A.~Strumia, JHEP {\bf 1312}, 089 (2013) [\href{https://arxiv.org/abs/1307.3536}{hep-ph/1307.3536}].

\bibitem{bezrukov2012} F.~Bezrukov, M.~Yu.~Kalmykov, B.~A.~Kniehl, and M.~Shaposhnikov, JHEP {\bf 1210}, 140 (2012) [\href{https://arxiv.org/abs/1205.2893}{hep-ph/1205.2893}].

\bibitem{Enqvist:2014bua} 
  K.~Enqvist, T.~Meriniemi and S.~Nurmi,
  JCAP {\bf 1407}, 025 (2014)
  [arXiv:1404.3699 [hep-ph]].

\bibitem{Markkanen:2017edu} 
  T.~Markkanen,
  arXiv:1711.07502 [gr-qc].


\bibitem{Enqvist:2014tta} 
  K.~Enqvist, S.~Nurmi and S.~Rusak,
  JCAP {\bf 1410}, no. 10, 064 (2014)
  doi:10.1088/1475-7516/2014/10/064
  [arXiv:1404.3631 [astro-ph.CO]].



\bibitem{Hardwick:2017fjo} 
  R.~J.~Hardwick, V.~Vennin, C.~T.~Byrnes, J.~Torrado and D.~Wands,
  JCAP {\bf 1710}, 018 (2017)
  [arXiv:1701.06473 [astro-ph.CO]].


\bibitem{Greene:1997fu}
      P.~B.~Greene, L.~Kofman, A.~D.~Linde, and A.~A.~Starobinsky,
      Phys.\ Rev.\ D {\bf 56}, 6175 (1997) [hep-ph/9705347].
  




\bibitem{felder1999} G.~Felder, L.~Kofman, and A.~Linde, Phys.\ Rev.\ D {\bf 59}, 123523 (1999) [\href{https://arxiv.org/abs/hep-ph/9812289v2}{hep-th/9812289}].

\bibitem{bezrukov2009} F.~Bezrukov, D.~Gorbunov, and M.~Shaposhnikov, JCAP {\bf 0906}, 029 (2009) [\href{https://arxiv.org/abs/0812.3622}{hep-ph/0812.3622}].

\bibitem{garciabellido2009} J.~Garcia-Bellido, D.~G.~Figueroa, and J.~Rubio, Phys.\ Rev.\ D {\bf 79}, 063531 (2009) [\href{https://arxiv.org/abs/0812.4624}{hep-ph/0812.4624}].

\bibitem{repond2016} Jo\"el Repond and Javier Rubio, JCAP {\bf 1607}, 043  (2016) [\href{https://arxiv.org/abs/1604.08238}{astro-ph/1604.08238}].

\bibitem{allahverdi2011} R.~Allahverdi, A.~Ferrantelli, J.~Garcia-Bellido, and A.~Mazumdar, Phys.\ Rev.\ D {\bf 83}, 123507 (2011) [\href{https://arxiv.org/abs/1103.2123}{hep-ph/1103.2123}].


\bibitem{McAllister:2008hb} 
 L.~McAllister, E.~Silverstein and A.~Westphal,
  Phys.\ Rev.\ D {\bf 82}, 046003 (2010)
  [arXiv:0808.0706 [hep-th]].
 




\bibitem{Adshead:2017xll} 
  P.~Adshead, J.~T.~Giblin and Z.~J.~Weiner,
  Phys.\ Rev.\ D {\bf 96}, no. 12, 123512 (2017)
  [arXiv:1708.02944 [hep-ph]].





\bibitem{Enqvist:2015sua} 
  K.~Enqvist, S.~Nurmi, S.~Rusak and D.~Weir,
  JCAP {\bf 1602}, no. 02, 057 (2016)
  doi:10.1088/1475-7516/2016/02/057
  [arXiv:1506.06895 [astro-ph.CO]].



\bibitem{Lozanov:2017hjm} 
  K.~D.~Lozanov and M.~A.~Amin,
 ``Self-resonance after inflation: oscillons, transients and radiation domination,''
  Phys.\ Rev.\ D {\bf 97}, no. 2, 023533 (2018)
  [arXiv:1710.06851 [astro-ph.CO]].

\bibitem{Amin:2010dc} 
  M.~A.~Amin, R.~Easther and H.~Finkel,
``Inflaton Fragmentation and Oscillon Formation in Three Dimensions,''
  JCAP {\bf 1012}, 001 (2010)
  [arXiv:1009.2505 [astro-ph.CO]].




\bibitem{Agashe:2014kda} 
 K.~A.~Olive {\it et al.} [Particle Data Group],
  Chin.\ Phys.\ C {\bf 38}, 090001 (2014).



\bibitem{Kolb:1990vq} 
  E.~W.~Kolb and M.~S.~Turner,
  Front.\ Phys.\  {\bf 69}, 1 (1990).







\bibitem{Giudice:2000ex} 
  G.~F.~Giudice, E.~W.~Kolb and A.~Riotto,
  Phys.\ Rev.\ D {\bf 64}, 023508 (2001)
  [hep-ph/0005123].


\bibitem{Chung:1998rq} 
D.~J.~H.~Chung, E.~W.~Kolb and A.~Riotto,
  Phys.\ Rev.\ D {\bf 60}, 063504 (1999)
  [hep-ph/9809453].






    \bibitem{Dvali:2003em} 
G.~Dvali, A.~Gruzinov and M.~Zaldarriaga,
  Phys.\ Rev.\ D {\bf 69}, 023505 (2004)
  [astro-ph/0303591].














  
\bibitem{Jimenez:2016ckl} 
  C.~P.~Garay, L.~Verde and R.~Jimenez,
  Phys.\ Dark Univ.\  {\bf 15}, 31 (2017)
  [arXiv:1602.08430 [astro-ph.CO]].


\bibitem{Palanque-Delabrouille:2015pga} 
  N.~Palanque-Delabrouille {\it et al.},
  JCAP {\bf 1511}, no. 11, 011 (2015)
  [arXiv:1506.05976 [astro-ph.CO]].

\bibitem{Cuesta:2015iho} 
  A.~J.~Cuesta, V.~Niro and L.~Verde,
  Phys.\ Dark Univ.\  {\bf 13}, 77 (2016)
  [arXiv:1511.05983 [astro-ph.CO]].

\bibitem{Hu:1997mj} 
  W.~Hu, D.~J.~Eisenstein and M.~Tegmark,
  Phys.\ Rev.\ Lett.\  {\bf 80}, 5255 (1998)
  [astro-ph/9712057].
  
  
  \bibitem{Lesgourgues:2006nd} 
  J.~Lesgourgues and S.~Pastor,
  Phys.\ Rept.\  {\bf 429}, 307 (2006)
  [astro-ph/0603494].
  
%




  \bibitem{Kusenko:2017kdr} 
  M.~Kawasaki, L.~Pearce, L.~Yang and A.~Kusenko,
  Phys.\ Rev.\ D {\bf 95}, no. 10, 103006 (2017)
  [arXiv:1701.02175 [hep-ph]].














\end{thebibliography}
\end{document}